\documentclass[twocolumn,times,tighten]{aastex631}
\usepackage{amsmath,amstext}

\newcommand{\rxj}{RXJ1007}
\newcommand{\rxjfull}{RX J100742.53+380046.6}
\newcommand\sloanr{r$^\prime$}
\newcommand\sloang{g$^\prime$}
\newcommand\sloangr{g$^\prime$ - r$^\prime$}
\newcommand\Msol{M$_{\odot}$}
\newcommand\kms{km s$^{-1}$}
\newcommand\slos{$\sigma_{los}$}

\newcommand{\masstwo}{$\mathrm{M}_{200}$}

\newcommand{\rtwo}{$\mathrm{r}_{200}$}
\newcommand{\hh}{$^{h}$}
\newcommand{\hm}{$^{m}$}

\usepackage[all]{hypcap}

\graphicspath{{./}{figures/}}

\begin{document}
\title{Independent Evidence for earlier formation epochs of fossil groups of galaxies through the intracluster light: the case for \rxjfull}

\author{Renato A. Dupke}
\affiliation{Eureka Scientific, 2452 Delmer St. Suite 100,Oakland, CA 94602, USA}
\affiliation{Observat\'orio Nacional, Rua General Jos\'e Cristino, 77, Bairro Imperial de S\~ao Crist\'ov\~ao, Rio de Janeiro, 20921-400, Brazil}
\affiliation{Department of Astronomy, University of Michigan, 311 West Hall, 1085 South University Ave., Ann Arbor, MI 48109-1107}
\affiliation{Department of Physics \& Astronomy, University of Alabama, Box 870324, Tuscaloosa, AL 35487, USA}

\author [0000-0002-6090-2853]{Jimenez-Teja, Yolanda}
\affiliation{Instituto de Astrofísica de Andalucía, Glorieta de la Astronomía s/n 18008 Granada, Spain}
\author{Su, Yuanyuan}
\affiliation{Department of Physics and Astronomy, University of Kentucky, 505 Rose street, Lexington, KY 40506, USA}
\author[0000-0002-7272-9234]{Carrasco, Eleazar R.}
\affiliation{Gemini Observatory/National Optical Infrared Research Laboratory, Casilla 603, La Serena, Chile}
\author[0000-0002-6610-2048]{Koekemoer, Anton M.}
\affiliation{Space Telescope Science Institute, 3700 San Martin Dr., Baltimore, MD 21218, USA}
\author[0000-0002-4949-8351]{Batalha, Rebeca}
\affiliation{Observat\'orio Nacional, Rua General Jos\'e Cristino, 77, Bairro Imperial de S\~ao Crist\'ov\~ao, Rio de Janeiro, 20921-400, Brazil}
\affiliation{Department of Physics \& Astronomy, University of Alabama, Box 870324, Tuscaloosa, AL 35487, USA}
\author{Johnson, Lucas}
\affiliation{Department of Physical Sciences, University of West Alabama, 100 US-11, Livingston, AL 35470}
\author{Irwin, Jimmy}
\affiliation{Department of Physics \& Astronomy, University of Alabama, Box 870324, Tuscaloosa, AL 35487, USA}
\author[0000-0002-3031-2326]{Miller, Eric}
\affiliation{Kavli Institute for Astrophysics and Space Research, Massachusetts Institute of Technology, 77 Massachusetts Avenue, Cambridge, MA 02139, USA}
\author{Dimauro, Paola}
\affiliation{Observat\'orio Nacional, Rua General Jos\'e Cristino, 77, Bairro Imperial de S\~ao Crist\'ov\~ao, Rio de Janeiro, 20921-400, Brazil}
\affiliation{INAF - Osservatorio Astronomico di Roma, Via di Frascati 33, 00078, Monte Porzio Catone, Italy}
\author{de Oliveira, N\'icolas O. L.}
\affiliation{Observat\'orio Nacional, Rua General Jos\'e Cristino, 77, Bairro Imperial de S\~ao Crist\'ov\~ao, Rio de Janeiro, 20921-400, Brazil}
\author{Vilchez, Jose}
\affiliation{Instituto de Astrofísica de Andalucía, Glorieta de la Astronomía s/n 18008 Granada, Spain}




\begin{abstract}
Fossil groups (FG) of galaxies still present a puzzle to theories of structure formation.  Despite the low number of bright galaxies, they have relatively high velocity dispersions and ICM temperatures often corresponding to cluster-like potential wells. Their measured concentrations are typically high, indicating early formation epochs as expected from the originally proposed scenario for their origin as being older undisturbed systems. This is, however, in contradiction with the typical lack of expected well developed cool cores. Here, we apply a cluster dynamical indicator recently discovered in the intracluster light fraction (ICLf) to a classic FG, RX J1000742.53+380046.6, to assess its dynamical state. We also refine that indicator to use as an independent age estimator. We find negative radial temperature and metal abundance gradients, the abundance achieving supersolar values at the hot core. The X-ray flux concentration is consistent with that of cool core systems. The ICLf analysis provides an independent probe of the system's dynamical state and shows that the system is very relaxed, more than all clusters, where the same analysis has been performed. The specific ICLf is more $\sim$5 times higher than any of the clusters previously analyzed, which is consistent with an older non-interactive galaxy system that had its last merging event within the last $\sim$5Gyr. The specific ICLf is predicted to be an important new tool to identify fossil systems and to constrain the relative age of clusters.

\end{abstract}


\section{Introduction} 

Fossil groups (FGs) are usually characterized as systems dominated by a single giant elliptical galaxy at least two magnitudes brighter than the second ranked galaxy 
($\Delta m_{1,2}\geq2$), within half the virial radius r$_{200}$,  \citep{jones2003} with extended X-ray emission corresponding to luminosities of more than $ 10^{42}$ h$_{50}^{-2}$ erg/s). Even though the first FG was discovered more than two decades ago \citep{ponman1994}, their origin and evolution are still debated. FGs were originally thought to be the cannibalistic remains of galaxy groups that lost energy through dynamical friction \citep[e.g.,][]{mulchaey1999}. 
Given the expected large times involved in dynamical friction and the lack of evidence of clear X-ray substructures, the original explanation for their nature was that FGs formed early and have been undisturbed for a very long time \citep[e.g.,][]{ponman1994,jones2003,vikhlinin1999}.  Hereafter, we refer to this scenario as the "standard model". 

Further X-ray and optical measurements have shown an increasing number of unusual characteristics in many FGs, which challenged the standard model. The intracluster gas temperatures (T$_X$) of FGs were often found to be similar to that of galaxy clusters, sometimes in excess of 4 keV  \citep[e.g.,][]{habib06,khosroshahi2007}. Measurements of galaxy velocity dispersion ($\sigma_{los}$) in FGs \citep[e.g.,][]{edu2006,proctor2011} were found to be consistent with the measured T$_X$  \citep[e.g.][]{khosroshahi2007,eric-sample}, indicating a mass range that encompasses that of medium sized clusters.  The lack of bright galaxies near the central bright cluster galaxy (BCG) in these cluster-sized systems makes them stand out.

The small number of combined detailed multi-wavelength X-ray and optical studies of FGs observed so far makes it difficult to provide an unambiguous answer about their formation epochs. On one hand, they seem to have high values of the concentration parameter, \textbf{c$_{200}$}, defined as the ratio r$_{200}$ to r$_s$, the NFW scaling radius \citep{khosroshahi2007}. Given the correlation found between c$_{200}$ and formation epoch in Cold Dark Matter ($\Lambda$CDM) simulations, the typical FG c$_{200}$ would be associated to earlier formation epochs \cite[e.g.,][]{risa,deason13}\footnote{Note that the definition of formation time is often the epoch at which 50\% of the system’s final virial mass is assembled, while \cite{risa2002} has a more ample definition independent of the specific parametrization}. On the other hand, the cooling time of FGs is observed to be usually significantly less than the Hubble time  \citep[e.g.,][]{ming2004,habib04,habib06,khosroshahi2006}, but they often lack the \textit{large} cool cores expected for a very old and undisturbed system and a paucity of central AGN activity \citep{miraghaei14,habib17}. Therefore, with currently available data, it is difficult to solve the above mentioned peculiarities of FGs and to determine whether they represent a ``physical'' class on their own.

Not all systems classified as FGs show the above mentioned paradoxical characteristics \citep[e.g.,][]{ming2009,fgcore,voevodkin2010}. This in part can be due to the ``purity'' of the samples selected for analysis.
Even though there is theoretical support for a relation between high magnitude gap and early accretion of a significant fraction of the system's mass  \citep[e.g.,][]{donghia2005}, $\Delta m_{1,2}$ alone is not a necessary condition for the system to be identified as an FG \citep[e.g.,][]{dariush}, since the majority of them will lose the magnitude gap in a few Gyr, i.e., the magnitude gap of an individual system is highly transitory \citep{kundert,benda2008,goza14}. Therefore, this selection criterion is prone to be affected by many physical and observational systematics. The infall of new galaxies by the group may reduce  the magnitude gap while the opposite may happen due to bright galaxies’ mergers with the BCG. The simple orbital motion of member galaxies will produce a variance in the gap that can be as high as 2 mags \citep{kundert,raouf2014}. Furthermore, this type of selection is susceptible to galaxy membership mis-identification and/or poor completeness \citep{voevodkin2010}. \cite{li2020} studied the evolution of the most stellar deficient groups in N-Body + semi-analytical cosmological simulations and found that low mass FGs ($10^{13.4} M_{\odot}<M<10^{13.6}M_{\odot}$) form earlier than normal groups and have lower stellar mass fraction. They also found that selections of FGs based on magnitude gaps show relatively low levels of purity. Nevertheless, magnitude gaps are the ``cheapest'' markers. Other, more physically supported, markers are expensive observation-wise. For example, as mentioned previously, concentration measurements such as c$_{200}$ are a good physical indicator of early formation epoch, but to measure it, one needs good X-ray or lensing observations as well.

Using a careful criteria to maximize purity we have selected a sample of FGs with good multi-wavelength data that can help to clarify the nature of FGs. The original selection started with the maxBCG cluster catalog \citep[][]{koster07}, where on top of a magnitude gap criteria, low richness--high X-ray luminosity systems were included. FGs tend to have relatively low richness with respect to their mass, as pointed out by \cite{rob}. Follow-up with Chandra snapshots allowed to restrict the sample further based on the absence of  cool cores and AGN activity, which were further confirmed by deeper XMM-Newton observations. We also obtained large-scale spectroscopy of the targets to verify membership. Additionally, we used  multi-wavelength HST observations, which, when coupled with the other high quality observations, can provide invaluable information about the intracluster light (\textbf{ICL}) distribution that can act as both a dynamical probe and age indicator, as explained below. Here we present the results for the first FG where this combined analysis has been applied.

Recent advances in measuring the ICL, i.e., stars that are not bound to individual galaxies in a cluster,  provided more robust techniques that allowed us to place constraints to the nature of FGs. For the sake of argument we can broadly separate the ICL growth processes in two types: those that increase together with the system’s mass (a), from those that do not (b). The first type includes cluster merging, where galaxy-galaxy interaction is significantly enhanced increasing disruption of dwarf galaxies and mergers with the BCGs. This increases the ICL production over the sum of the pre-merging individual sub-cluster's ICL \citep[e.g.,  ][]{jesse07,rudick11,yoli-whl}\footnote{Notice that CICLE is not subject to the systematics involved in measuring ICL thorough surface brightness cut-off methods. So in Figure 2 of \cite{rudick11} one would not expect to see the ICLf "dips" during mergers.}. Type (b) would include regular tidal stripping due to internal dynamical friction. Other minor processes such as star formation of stripped gas  \citep[e.g.][]{ming2010} can be included in (a) or (b).

\cite{yoli-clash} applied a  sophisticated ICL measuring technique to a sub-sample of 10 galaxy clusters from the Cluster Lensing And Supernova survey with Hubble (CLASH) survey and the Bullet cluster. Their results have been further expanded to include other HST Frontier Field clusters \citep{nicolas}, RELICS clusters \citep{yoli-relics} and even ground data using the Javalambre Photometric Local Universe Survey (J-PLUS) data \citep{yoli-coma}. We hereafter refer to the HST-observed sample mentioned above as \textbf{ICLHST}. Their results showed that merging clusters had higher intracluster light fraction (\textbf{ICLf}) fluxes than the relaxed ones and, more interestingly, that merging (dynamically active) clusters had an excess flux in the 4000-5000 \AA~rest-frame wavelength range, which corresponds roughly with the peak emission of main sequence stars of late-A to early-F spectral types. The ICLf is defined as the ratio of ICL to total (i.e., ICL+galaxies) fluxes.
They hypothesized that mergers violently amplify tidal stripping, rapidly removing stars from the outer part of galaxies. Furthermore, these merger-induced stripped and relatively short-lived bright stars would temporarily increase the 4000-5000 \AA~ICL flux before evolving out of the main sequence, reducing their flux contribution and returning the ICLf wavelength distribution to that characteristic of relaxed cluster, i.e., flat. So, even though the full ICLf at any single time may not be a robust indicator of the cluster’s dynamical state, as mentioned in \cite{rudick11}, \textit{the ICLf in particular wavelength ranges can provide information about the cluster’s dynamical state at a particular time.}

It is expected that the ICL continuously increases with cluster mass (e.g., \cite{lin2004,rudick11,zhang2011}). The heuristic classification of ICL growth types described in the previous paragraphs suggests that the ICL production occurs in \textit{two} different regimes: a \textit{steady regime}, where the ICL is produced by tidal stripping enhanced by internal dynamical friction (in between merging events) and a \textit{violent regime} during mergers with a fast injection of stars into the ICL. That would imply that the growth over time of the system's mass and that of ICL are different.
 In this very simplified view, when two clusters merge, the final mass is the sum of their pre-merging masses but the ICL of the merged system would be greater than the sum of the ICL of each pre-merging system, given that the merging process itself would produce new ICL through the violent regime.
In between mergers, the ICL would still continue to grow even though the cluster's mass would not. 
So, if two clusters formed at the same time, following  different merging trees and keeping similar average merging histories, i.e., similar overall mass distributions of constituent halos and subhalos, one would expect them to roughly have similar masses and ICL-to-mass ratios at any particular epoch. The latter would steadily grow even after all surrounding bound halos collapse into the final system.  
On the other hand, if one of these systems (S1) formed earlier than the other (S2), it would achieve its maximal final mass prior to S2 and start growing its ICL-to-mass ratio under the steady regime only. So, at the moment S2 had its last merger, S1 would have the same mass but a higher ICL-to-mass ratio than S2. FGs, within the standard model, would correspond to S1 and one would thus expect them to have an enhanced ICL-to-mass ratio. The same reasoning can be applied to the ICLf-to-Mass ratio.

Here, we use this prediction to probe the age and the ubiquity of the standard scenario for a ``classic'' FG SDSS J100742.53+3800466 \footnote {Its MAST name is \rxjfull~ and its XMM-Newton Science Archive is RX J1007+3800}, RA=10 07 42.5, Dec=+38$^{\circ}$ 00 47.5 at z$\sim$0.112, which we denote throughout this paper as \rxj. We use the standard $\Lambda$CDM cosmology with $\Omega_m$ = 0.286, $\Omega_{\Lambda}$ = 0.714, and $H_0$ = 69.6~ km~$s^{-1}$ Mpc$^{-1}$, unless stated otherwise.  Uncertainties are 1-$\sigma$ and upper/lower limits are 3-$\sigma$.

\section{Data: observations, calibration and methodology} \label{sec_data}
\subsection{HST}
Three orbits of the Hubble Space Telescope (HST) time were allocated to image the fossil group RX J1007+3800, in  Cycle 25 (PI: Dupke). This proposal was part of a larger program, which also granted 
53 ksec XMM-Newton (XMM) observation. HST images were taken with the Advanced Camera for Surveys (ACS) on May 11th 2018, using two different filters: F435W (one orbit) and F606W (two orbits).  Each orbit was divided into four dithered exposures in order to cover the gap between the two ACS detectors, mitigate against bad detector pixels, to enable cosmic ray rejection, and to provide sub-pixel sampling in order to improve the final resolution of the stacked images. Individual raw exposures were first processed by the default HST pipeline in MAST with CALACS\footnote{http://www.stsci.edu/hst/acs/performance/calacs\_cte/calacs\_cte.html}, which applies several detector-level calibrations including bias subtraction, flat-fielding, and correction for charge transfer efficiency losses, as well as applying geometric distortion corrections using drizzlepac\footnote{https://drizzlepac.readthedocs.io/en/latest/}. We then applied additional processing to these calibrated exposures, in particular improving the rejection of cosmic rays and bad pixels, and significant improvements to the astrometric alignment, following procedures first developed by \cite{koekemoer2011}, achieving better astrometric precision than provided by the default pipeline. We thereby produced combined mosaics with parameters optimized to our specific observational design, in order to improve the quality of the final mosaics, obtaining cleaner stacked images and a better sampling of the PSF, to take advantage of our sub-pixel dithering observing strategy. Our final mosaics are at a pixel scale of $0.03^{\prime\prime}$ per pixel for optimal PSF sampling, as well as $0.06^{\prime\prime}$ per pixel for computational-intensive analysis of the ICL. Final exposure depths were 2255 and 4493 seconds for the F435W and F606W filters respectively.\\

We measured the ICL of both images using CICLE \citep[CHEFs ICL Estimator,][]{yoli-pandora}. CICLE is an algorithm specially designed to estimate the ICLf in galaxy clusters. In order to disentangle the galactic contribution from that of the ICL, foreground stars are usually masked out and galaxies are fitted using mathematical orthonormal bases called CHEFs \citep[Chebyshev-Fourier functions,][]{yoli-chef}. CHEFs efficiency and flexibility to model the surface distribution of galaxies are directly inherited from the mathematical properties of the Chebyshev rational functions and Fourier modes that compose them. Moreover, the CHEFs have proven to be able to fit very different galaxy morphologies \citep{yoli-chef}, including BCGs \citep{zitrin2010}. This flexibility allows the CHEFs to fit any regular galaxy of a cluster since they appear as reasonably well defined clumps of luminosity over the smooth, extended ICL background. However, the BCG extended halo can be easily misidentified with ICL since the transition from the BCG-dominated to the ICL-dominated area is gradual. For this reason CICLE outlines the limits of the BCG applying a more sophisticated approach before modeling it out with the CHEFs. Basically, CICLE calculates the curvature at each point of the BCG+ICL composite surface, intuitively understood as 
the intensity surface profile. The curvature is deﬁned as the difference in the slope between a point and its surroundings. Once the region where the slope of the composite surfaces is maximized the transition from BCG to the ambient ICL is determined. Then,  CICLE \textbf{removes} the BCG component and uses the surrounding region to interpolate the underlying ICL. To use the maximum slope change as the transition from BCG to the “true” ICL is physically justified because the stars from the BCG are supposed to have different kinematics as those from the ICL, since they have originated through different processes.
CICLE is based on the assumption that the two profiles, the BCG and the ICL, have a different inclination, so the points where the curvature is maximized indicate the transition from the BCG halo to the ICL. This algorithm has been successfully tested with mock data \citep{yoli-pandora}, space and ground-based observations to both nearby \citep{yoli-coma} and intermediate-redshift clusters \citep{yoli-clash,yoli-whl,nicolas}. Once the ICL maps are computed, we calculate the ICLf.

\subsection{Gemini}\label{sec:gemdata}

\begin{figure}[t!]
\includegraphics[width=\hsize]{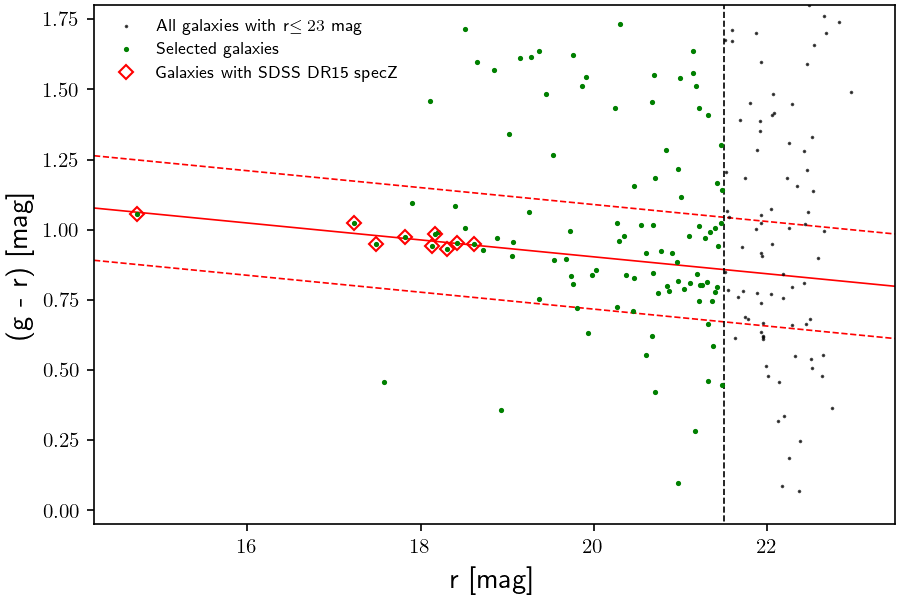}
\caption{Color-magnitude diagram of all object classified as galaxies inside the GMOS field of view. Gray points represent all galaxies with SDSS DR15 magnitudes brighter than \sloanr$=23$ mag and the green dots represent the selected FG galaxies candidates. About $\sim 61$\% of the selected sample (66 galaxies) were included in two GMOS masks (see text).}
\label{cmdsel}
\end{figure}

In previous works, we used to identify the member candidates using spectroscopic redshifts \citep{yoli-pandora,yoli-clash,yoli-coma}, which have the advantage of being much more precise than photometric redshifts. On the other hand, spectroscopic samples usually are not complete, and FGs have a paucity of bright galaxies in the central regions so that mis-identifications errors can bias the results significantly. So, we prioritized purity over completeness. In the case of RX J1007+3800, only 9 galaxies with spectroscopic redshift associated to the peak of the cluster at $z\sim0.112$ (within $\pm 1500$ \kms) and inside an area of $\sim 0.61 \times 0.61$ Mpc$^{2}$ ($\sim 6 \times 6$ arcmin$^2$) are publicly available \citep{aguado2019}. The number of galaxies associated to the cluster increases to 16 when the search area is extended $\sim 1.4 \times 1.4$ Mpc$^{2}$ ($\sim 12 \times 12$ arcmin$^2$). To overcome the lack of spectral information in the FG field, in particular in the central regions, where there are only a few galaxies with spectroscopic redshift information, new spectroscopic redshifts were obtained using the Gemini Multi-Object Spectrograph mounted at Gemini North telescope \citep[GMOS, ][]{hook2004}. We used the SDSS DR15 to construct the color-magnitude diagram (CMD) and select the galaxies for spectroscopic follow-up inside the GMOS field of view ($5.5 \times 5.5$ arcmin$^2$). Figure \ref{cmdsel} shows the CMD of all galaxies with SDSS photometry and \sloanr$\le 23$ mag (gray points). One-hundred and eight galaxies brighter than \sloanr$=21.5$ were selected as member candidates for spectroscopy (green dots in Fig. \ref{cmdsel}). A total of 66 galaxies ($\sim 61$\% of the selected sample) were included in two GMOS masks, prioritizing those galaxies located within $\pm 1.5\sigma$ around the best-fit for the galaxies lying in the red sequence (red solid and dashed lines in Fig. \ref{cmdsel}). It is worth noting that the apparent magnitude cutoff (\sloanr$=21.5$ mag) is one magnitude deeper than the second ranked galaxy ($\sim M^{*}_{r} + 3.5$ mag).

The two GMOS masks were observed at different nights under the Program ID: GN-2019A-FT-206 (PI: Dupke). Mask 1 was observed on 2019 May 25 UT during dark time, under clear sky and poor seeing ($\sim 1$\arcsec) conditions. The second mask was observed a month later, on 2019 June 20 UT, during dark time, with some cirrus (patchy cloudy) and under good seeing ($\sim$ 0\farcs7) conditions. The spectra were acquired using the R400 grating centered at 6250\AA, using 1\arcsec\ slitest and $2 \times 2$ binning. Wavelength offsets of 100\AA\ toward the blue and the red were applied between exposures to cover the gaps between CCDs. At each wavelength setting, spectroscopic flats and CuAr comparison lamps spectra were taken before or after each science exposure. The science spectra were flux-calibrated using the spectrophotometric standard star Feige 34 observed with the same instrument setup than the science images, but on a different night (April 24, 2019 UT) and under different observing conditions. Therefore, only a relative flux calibration of the science spectra are provided.

The science images were reduced using the Gemini GMOS package version 1.14\footnote{ https://www.gemini.edu/observing/phase-iii/understanding-and-processing-data/data-processing-software/gemini-iraf-general} following the standard procedures for multi-object spectroscopic (MOS) observations. All science and calibration exposures were over-scanned, bias-subtracted and trimmed. The two-dimensional science exposures were then flat-fielded, wavelength-calibrated, distortion-corrected, and extracted to one-dimensional format. The final wavelength solution has an average $\textit{rms}$ of $\sim0.25$ \AA. The resolution of the one-dimensional extracted spectra is $\sim 7.1$\AA\ (measured from the sky lines FWHM), with a dispersion of $\sim 1.5$ \AA~ pixel$^{-1}$, and covering a wavelength interval between $\sim 4000$~\AA~ -- $\sim 9200$~\AA.

The redshifts of the selected galaxies in the FG field were determined using the {\tt RV} package inside IRAF. All spectra were cross-correlated with four high signal-to-noise (S/N) templates using the program {\tt FXCOR}. The redshift errors were estimated based on the \textit{R} statistic value of \citet{tonry1979}. For galaxies with obvious emission lines, a line-by-line Gaussian fit was employed using the routine {\tt RVIDLINES}. The errors of the measurements were estimated using the residual of the average redshift shifts of all measurements provided by the program. We were able to determine the redshifts for all 66 galaxies included in the two GMOS masks (100\% success rate) plus one additional galaxy found by chance in one of the slits. 

The GMOS field of view covers a physical area of 0.68 $\times$ 0.68 Mpc$^2$, corresponding roughly to the core of \rxj. To obtain a robust estimation of the dynamical parameters, we have to increase the number of member galaxies beyond the GMOS field of view. We used the SDSSDR15 database to retrieve the spectroscopic redshift information of all galaxies within a field of view  24\arcmin$\times$24\arcmin~($\sim 2.95 \times 2.95$ Mpc$^{2}$). The search was limited to galaxies with $0.0 < z < 0.3$. Our GMOS sample is then supplemented with additional 53 galaxies inside the above area and redshift interval. Of the 53 galaxy redshifts retrieved from SDSSDR15, 10 galaxies have redshifts estimation in common with GMOS. The redshifts obtained with GMOS agree well with those redshifts in SDSSDR15 database, with a mean difference of $42\pm20$ \kms (\textit{rms} of $63$ \kms) between both data sets. 

The final galaxy catalog contains 98 galaxies with secure redshift determinations. The average redshift, the one dimensional line-of-sight velocity dispersion and the number of member galaxies of the cluster were estimated using the robust bi-weight estimators of central location ($C_{BI}$) and scale ($S_{BI}$) of \citet{beers1990}, using an iterative procedure and applying a 3-$\sigma$ clipping algorithm to remove outliers. The best estimates of the location ($\overline Z$) and scale (\slos), as well as the number of member galaxies, the \rtwo~and the \masstwo~(see below) are shown in Table \ref{tab:rxjdympar}. The list of  member galaxies of \rxj, magnitudes, colors and  redshifts are shown in Table \ref{tab:galmem} in the Appendix \ref{appx:galaxy_redshift}. For completeness, Table \ref{tab:galnomem} in Appendix \ref{appx:galaxy_redshift} shows the catalog of the foreground and background galaxies observed with GMOS.

The upper panel of Figure \ref{histcmd} shows the redshift distribution of all galaxies with $z\le0.3$ within the 24\arcmin$\times$24\arcmin field of view. The red histogram shows the distribution of the 46 spectroscopic confirmed member galaxies of \rxj. The inset in the upper panel shows projected phase-space diagram for all spectroscopic confirmed member galaxies, given by the peculiar line-of-sight velocity of each member galaxy with respect to the mean cluster velocity, normalized by the velocity dispersion of the cluster (equation \ref{eq:pecvel}) as a function of the projected distance from the center of the cluster, normalized by the \rtwo. 

\begin{equation}
\frac{\Delta V_{i}}{\sigma} =  \frac{(z_{i} - \overline{z})}{(1 + \overline{z})} \frac{\rm{c}}{\sigma}
\label{eq:pecvel} 
\end{equation}

\begin{figure}[t!]
\includegraphics[width=\hsize]{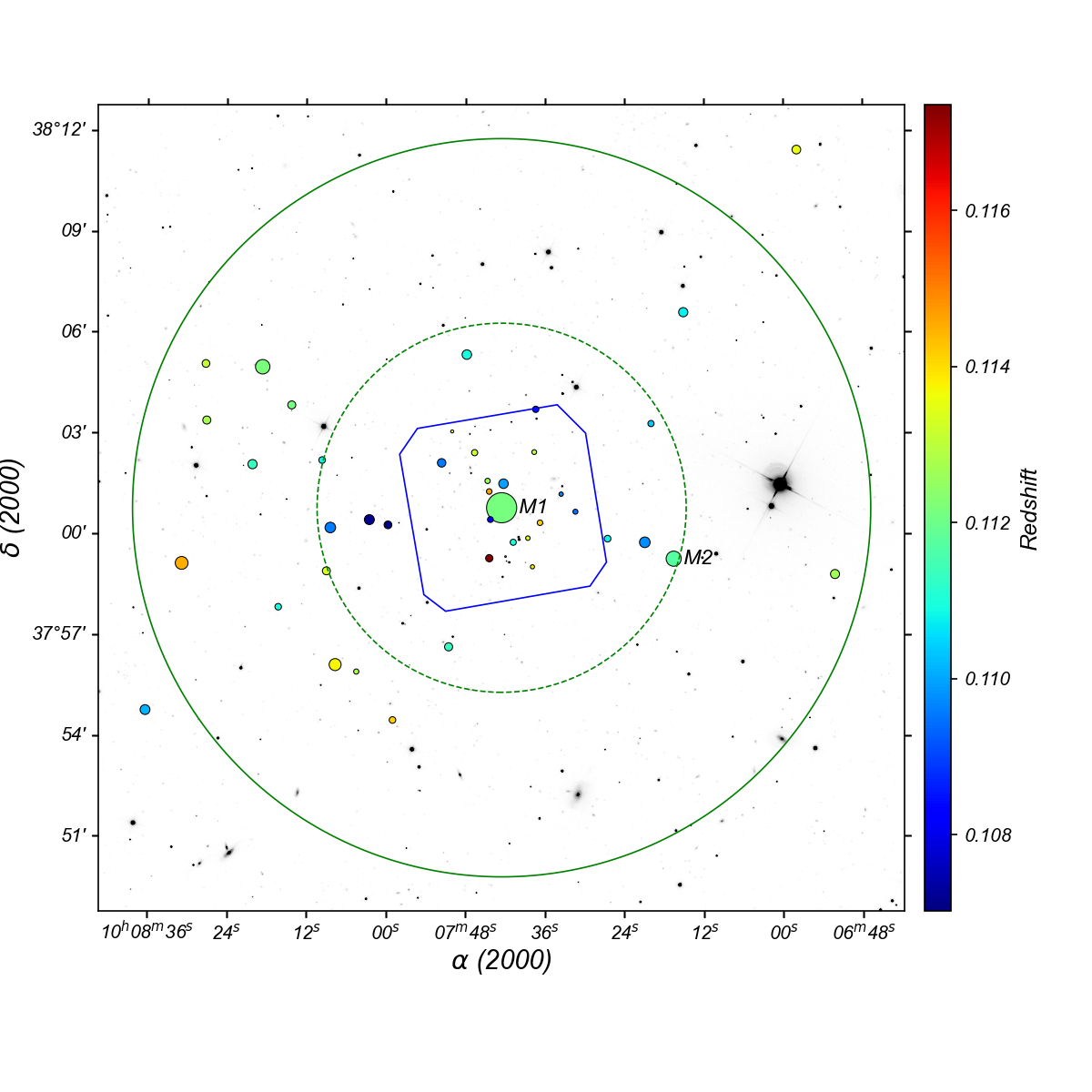}\\
\caption{Spatial distribution of the 46 spectroscopic confirmed member galaxies of \rxj. The circle sizes re proportional to the luminosity. THe distance from the first ranked galaxy (rSDSS = 14.73 AB mag) to the second (rSDSS = 17.18 AB mag) is 0.66 Mpc (dashed circle).}

\label{spatd}
\end{figure}

\begin{figure}[t!]
\includegraphics[width=\hsize]{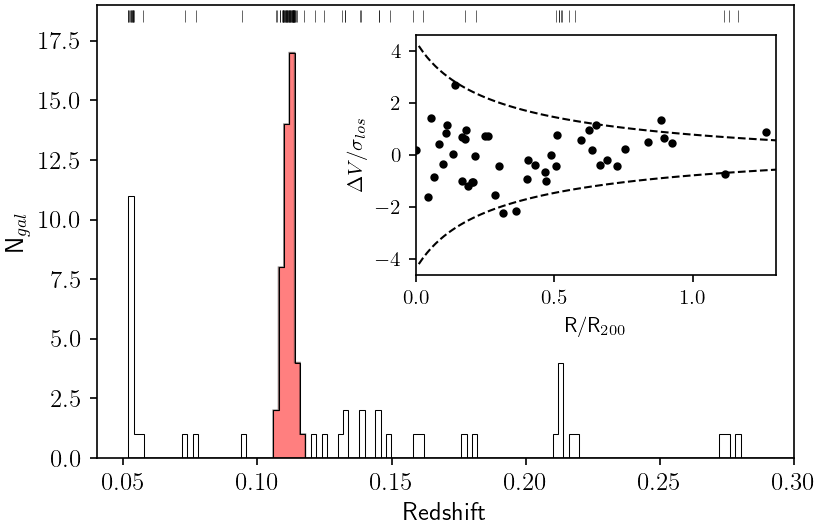}\\
\includegraphics[width=\hsize]{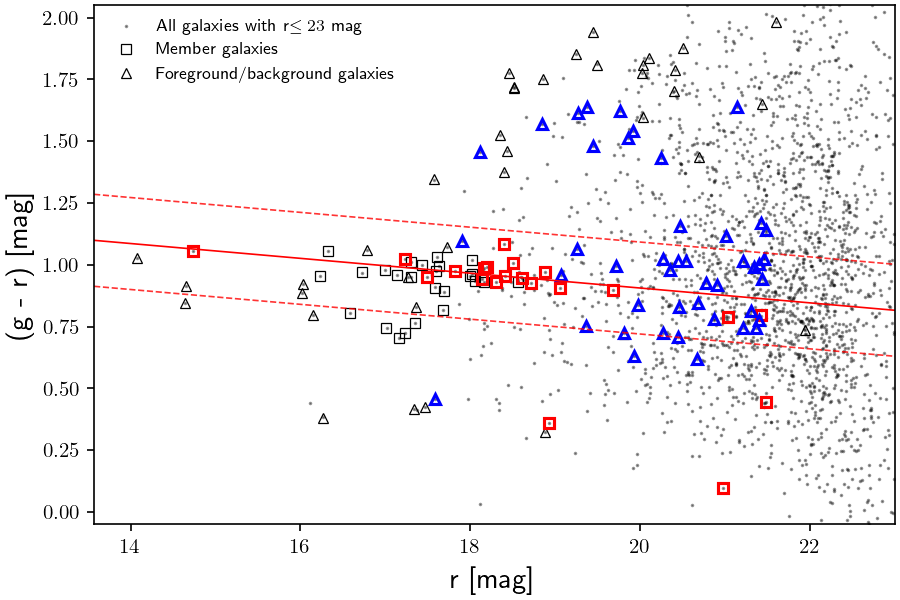}
\caption{\textit{Top Panel:} Redshift distribution of all galaxies with confirmed spectroscopic redshifts $\le0.3$ within a field of view of 24\arcmin$\times$24\arcmin. The red region shows the distribution of the 46 member galaxies of \rxj. The redshift of individual galaxies are represented by tick marks at the top of the histogram. The inset shows the phase space diagram normalized by the velocity dispersion as a function of the normalized radius. The dashed lines in the inset represent the  escape velocity for a NFW halo projected in the line of sight (see text). \textit{Bottom Panel:} Colour-magnitude diagram of all galaxies brighter than \sloanr=$23$ mag inside 24\arcmin$\times$24\arcmin in the \rxj~field (gray points). The squares and triangles represent the members and foreground/background galaxies, respectively. The galaxies observed with GMOS are represented by red squares (member galaxies) and blue triangles (foreground/background galaxies). The solid red line fits the red sequence of the FG, while the dashed lines represent the $\pm 1.5\sigma$ from the best-fit of the red sequence.}
\label{histcmd}
\end{figure}

The dashed lines in the inset represent the escape velocity for an NFW halo with the corresponding M$_{200}$ projected in the line of sight~ \citep{nfw1996,jaffe2015}. The galaxies located inside the dashed lines
are expected to be gravitationally bound to the cluster, i.e., located in the virialized region of \rxj. The spatial distribution of the spectroscopic confirmed members is shown in Figure \ref{spatd}. M1 an M2 are the first and second rank members with a $\Delta m_{1,2}=2.45$ and a projected separation of 0.66 Mpc.

The \masstwo~in Table \ref{tab:rxjdympar} was computed using the $\sigma$-\masstwo~scaling relation of \citet{munari2013} obtained from zoomed-in hydro-dynamical simulations of Dark Matter (DM) halos calibrated using Dark Matter particles and taking into account prescriptions for cooling, star formation, and Active Galactic Nuclei (AGN) feedback:

\begin{equation}
\sigma_{1D} = A_{1D}  \left[\frac{h(z)\,M_{200}}{10^{15} M_{\odot}}\right]^{\alpha}
\label{eq1}
\end{equation}

\noindent where $\sigma_{1D}$ is the one-dimensional (1D) velocity dispersion,
$A_{1D}=1177\pm4.2$ \kms, $\alpha = 0.364\pm0.002$, $h(z) = H(z)/H_{0}$
\kms~Mpc$^{-1}$, and \masstwo~ is the mass within \rtwo\footnote{\rtwo~is the
radius where the over-density is 200 times the critical density of the universe
and is defined as \rtwo$= (\sqrt{3}~\sigma_{\rm los})/ (10~ H(z))$
\citep{carlberg1997}}.
\begin{deluxetable}{lc}
\tabletypesize{\scriptsize}
\tablewidth{0pc}
\tablecolumns{2}
\tablecaption{Dynamical parameters\label{tab:rxjdympar}}
\tablehead{\colhead{Parameter} & \colhead {Values}}
\startdata
Cluster Center & 10\hh07\hm42\fs53, $+$38\degr00\arcmin46\farcs49 \\
$\overline{Z}$  & $0.111834\pm0.000480$ \\
N$_{mem}$ & 46 \\
\slos~(\kms) & $570\pm56$ \\
\rtwo~(Mpc) & $1.35\pm0.13$   \\
\masstwo ($10^{14}$ ~\Msol) &  $1.30\pm0.35$  \\
\enddata
\end{deluxetable}

\subsection{XMM} \label{sec:data_reduction}
\rxj$\;$was observed with XMM-Newton in 2010 and 2018 (PI: Dupke). Resulting effective exposures after cleaning flare contaminated time intervals are shown in Table \ref{tab:observations}. Only data from the European Photon Imaging Camera (EPIC) (MOS and PN detectors) are processed and reported in this paper. The standard Science Analysis System (SAS 18.0.0) pipeline tools were used throughout this analysis. SAS tools \textit{emchain}, \textit{epchain} for regular and also Out Of Time (OOT) events were used to generate calibrated event files from the raw data. \textit{mos-filter} and \textit{pn-filter} were subsequently used to remove the soft proton flares.
Point sources resolved with SAS tool {\it cheese} were removed and additional sources were subsequently removed manually using the {\it xmmselect} GUI after verification by eye using regions large enough to encompass at least one XMM PSF (20$^{\prime\prime}$) . 

The background is composed of several components, including soft proton flares (\textbf{SP}),  cosmic ray induced instrumental fluorescent lines, solar wind charge exchange (\textbf{SWCX}) and several astrophysical components, including the Local Bubble, the Galaxy Halo in the soft-bands and a power-law component for the cosmic X-ray background (\textbf{CXB}) (for details see www.cosmos.esa.int/web/xmm-newton/epic-background-components). We determined the main parameters of these components directly through spectral fittings following the recommendations of Snowden \& Kuntz (2014)\footnote{heasarc.gsfc.nasa.gov/docs/xmm/esas/cookbook/xmm-esas.html\#\\tex2html33}.

MOS and PN instrumental backgrounds were calibrated with Filter Wheel Closed (FWC) Data (as described below).
MOS 1 CCD\#6 in Observation 0824910201, CCDs \#3, \#4\& \#6 were not considered in Observations 0824910201 and 0824910101, due to known primary or secondary damage from micrometeorite hits.
The effective exposure times for the source data are shown in Table {\ref{tab:observations}.\\

\begin{table}[!h]
    \centering
    \caption{Cleaned effective exposures for each detector in all three observations.}
\begin{tabular}{ |p{1.5 cm}||p{1.5cm}|p{1.5cm}|p{1.5cm}|  }
\hline
 \multicolumn{4}{|c|}{Effective Exposure (ksec) } \\
 \hline
 & OBSID 0653450201 Nov. 2010 &OBSID 0824910101 May 2018 &OBSID 0824910201 Oct. 2018\\
 \hline
 MOS 1&29.5&23.4& 22.1\\
 MOS 2&29.7&22.7&22.9\\
 PN&28&46.7&19.3\\
  \hline   
  \end{tabular}
    \label{tab:observations}
\end{table}

Production of intermediate files necessary to create model background spectra for the specific regions analyzed here, including spectra and responses,  was made with the tools \textit{mos-spectra }and \textit{pn-spectra}, using the most recent Filter Wheel Closed (\textbf{FWC}) calibration data. Quiescent particle background (\textbf{QPB}) spectral models were generated with the tool \textit{mos\_back }and\textit{ pn\_back} with OOT events subtracted and the solid angle of the individual regions was derived from the task \textit{proton\_scale}.

\subsubsection{Background Treatment} \label{sec:bkg_treatment}
Given the complexity of XMM detector's background, we determined  the remaining non-quiescent and cosmic background through spectral fitting modeling all components, together with an absorbed CIE model for the intracluster gas residual emission for a large external annular region from 3$^{\prime}$--12$^{\prime}$ ($\sim$360-1,450) kpc, which starts at $\ge 0.5~r_{500}$) centered at \rxj, called hereafter the "outer region". We chose the outer region to be thick enough to have better statistics to constraint the other modelled background components, even at the cost of having to insert a source component (the FG emission), without overwhelming its contribution. Given that the system is at an intermediate redshift, we believe this is a good option and allows us to have slightly better statistics overall, given that we do not refit the slopes of the soft proton background for the each annulus. Spectra from all detectors were individually grouped to have a minimum of 20 counts per channel with the ftool \textit{grppha}. 

The FG contamination in this external region was modeled with an absorbed \textit{apec} model. The nH column (1.35$\times10^{20} cm^{-2}$) was chosen from the HI4PI Map  \citep{HI4PI} through the HEASARC nH tool\footnote{heasarc.gsfc.nasa.gov/cgi-bin/Tools/w3nh/w3nh.pl}.  Abundances listed here are with respect to the photospheric value \citep{angr}. Following the prescription laid out by \cite{snowden2004}, we introduced Gaussian components in the spectral model that were not included in the QBP spectra with energies of 1.486 keV (Al K$\alpha$) for MOS \& PN, 1.74 keV (Si K$\alpha$) for MOS to account for instrumental fluorescence lines and also five more Gaussians for the PN for the main contribution from Cu, Ni \& Zn at 7.49 keV, 7.11 keV, 8.05 keV, 8.62 keV, and 8.90 keV. Two extra Gaussians are included to account for potential SWCX strongest contribution to the O VII \& O VIII lines \citep[e.g.,][]{kuntz2019} at 0.56 and 0.65 keV.
Two more apec models were used, one for the Local Bubble emission with a fixed temperature of 0.1 keV and an absorbed one to account for the hot Galactic Halo with a fixed temperature of 0.26 keV. Their metal abundance and redshift are fixed at 1 and 0 respectively and their model normalizations were allowed to vary independently. The redshift of the FG emission was fixed at the nominal value. 
An absorbed power law with a slope of 1.46 was also incorporated to account for the CXB.  
 
To constrain the cosmic background component we added a  fourth data group set with ROSAT All-Sky Survey (RASS) spectrum of the off-source (1$^{\circ}$-2$^{\circ}$) background as a good approximation for the cosmic background in the direction of the target, obtained from the HEASARC\textit{ X-ray Background Tool}\footnote{heasarc.gsfc.nasa.gov/cgi-bin/Tools/xraybg/xraybg.pl} with the respective responses.
 
We then fit the background spectra with XSPEC 12.11.0 using the following model: gauss + gauss + gauss + gauss + gauss + gauss + cons*cons *(gauss + gauss + apec + (apec + apec + pow)*wabs + apec*wabs). The constant components are for the relative normalization between the detectors and the solid angle scaling.

 To determine the contamination level by soft protons (\textbf{SP}) a broken power law  with a break at 3.0 keV was added as a separate model using the diagonal responses included in the updated Current Calibration Files (\textbf{CCF}). The best fit values for the broken powerlaw slopes and the FG residual contamination in the outer region are shown in Table \ref{bestfit}. These slopes are fixed in the spectral fits for all internal regions.
 
 For the sake of comparison, we also carried out an alternative background strategy, in which we reduced the size of the outer region to a very external annulus, where no source contribution was expected, and added a carefully chosen external pointing to fix some of the CXB parameters. In this strategy we re-modelled the full background (including the QPB) in every cluster region. The results of this alternative method are mostly consistent with the the previous and are shown in Appendix \ref{appx:bkg_modeling}.\\
 
\section{Results}
\subsection{ICL}
\begin{figure*}
\label{ICLim}
\centering
\includegraphics[width=\hsize]{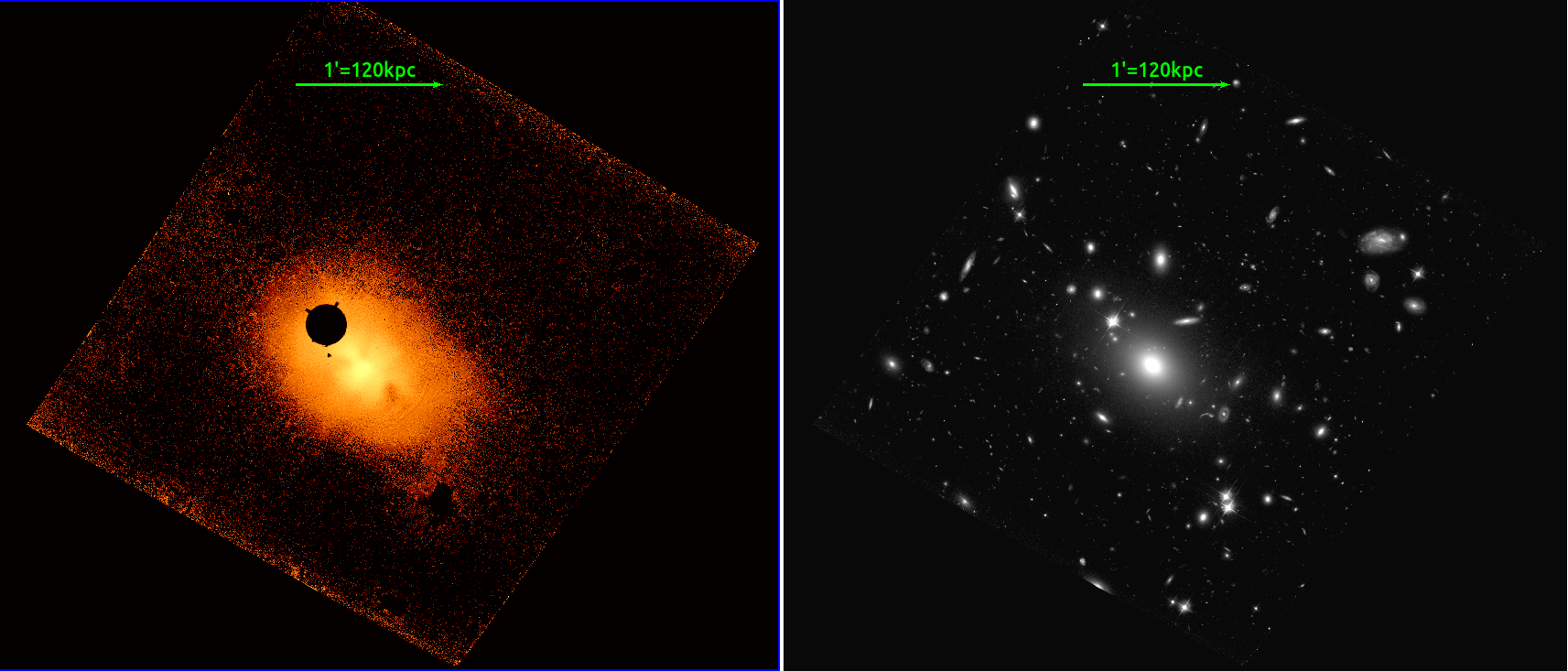}
\includegraphics[width=\hsize]{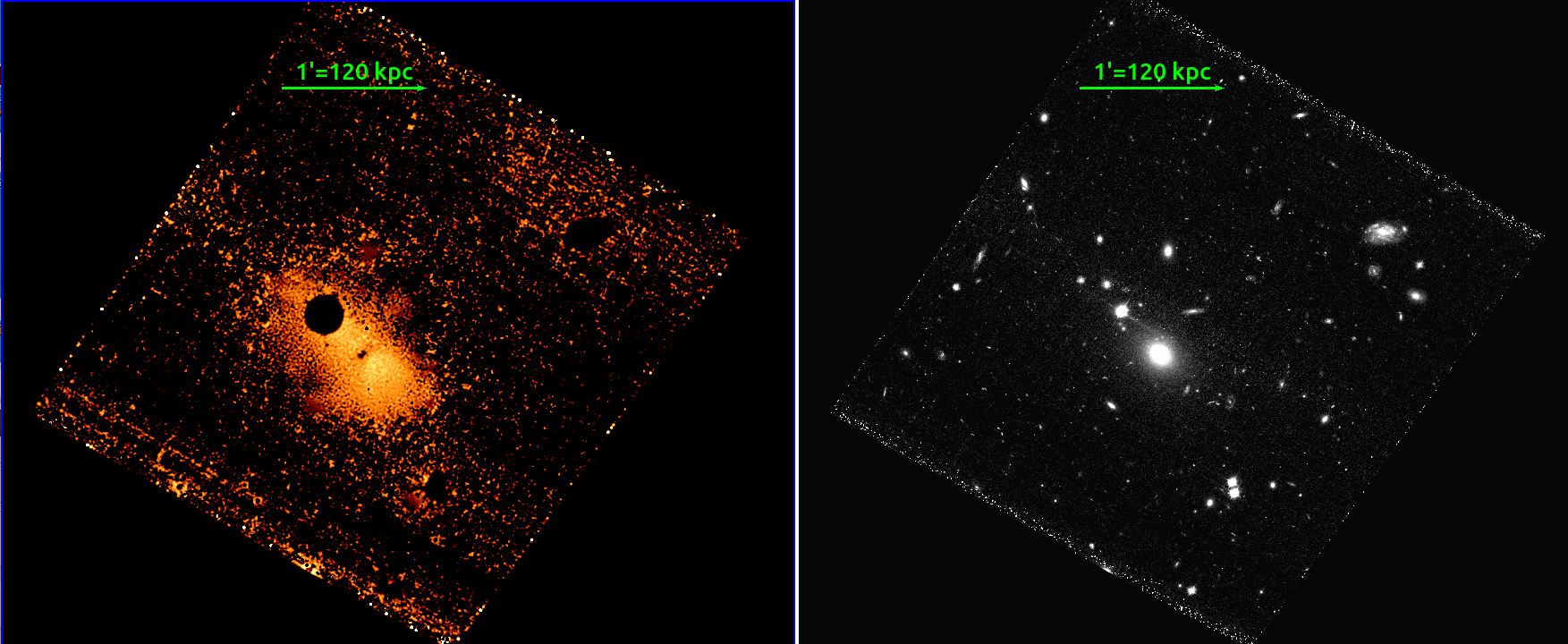}

\caption{CICLE ICL (left) and Full (right) images of \rxj, in the F606W filter (top) and F435W filter (bottom). The images are at a different brightness scale to improve visibility. North is up. East is left.
}
\end{figure*}
\begin{figure}
\begin{center}

\hspace*{-1.7cm}
\resizebox{11cm}{!} {\includegraphics{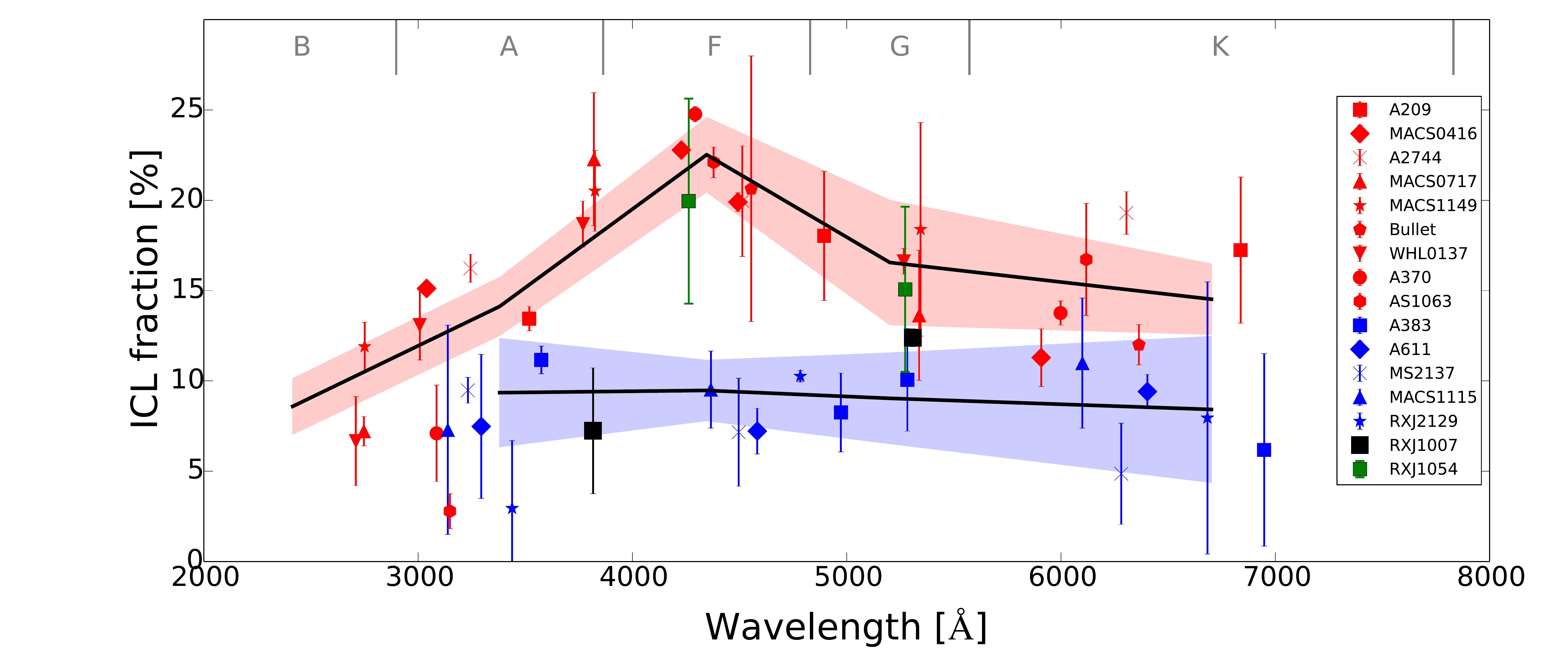}}
\caption{ICLf rest-frame color distribution for merging (red) and relaxed (blue) clusters studied in \citet{yoli-clash}. Black lines indicate the error weighted mean of each main sequence spectral-type subsample and colored shadowed areas, the mean of the errors. We also plot the values for a candidate fossil cluster RX JJ105452.03+552112.5 in green \cite{yoo}. Gray vertical lines at the top of the figure split the wavelength range expected for the peak emission for an average main sequence star, labeled with gray letters. Although the distribution for relaxed clusters is mostly flat, that of merging clusters shows an excess in the region corresponding to the emission peaks of late A- and early F-type stars. \rxj~ distribution is shown by the black squares and does not follow the typical behavior of a merging nor of a normal relaxed cluster. For the sake of visibility, we used a symbol size larger than the small errorbars for the F606W filter}
\end{center}
\label{all_clusters}
\end{figure}

CICLE was applied to the fully-calibrated images of \rxj, both in the F435W and the F606W bands. Results show a relatively compact ICL (especially, in the F606W filter), highly concentrated around the BCG (Figure \ref{ICLim}). Its distribution is smooth without signs of substructure, bumps or irregularities of significance. The obtained ICLfs are: $7.24\pm 3.48\%$ and $12.39\pm0.5\%$, for the F435W and F606W bands and detection limits of 137.7 kpc and 178.0 kpc, respectively. The error bars do \textit{include} both the photometric error associated to the measurement of the flux of the ICL and the cluster galaxies, and the theoretical error associated to the CICLE algorithm.
This latter error is calculated simulating images with the same observational and geometrical characteristics as the original images, (i.e., simulating a composite BCG+ICL surface using exponential profiles with the same effective radii and magnitudes as the original objects, and adding noise with the same signal-to-noise as our HST data). CICLE is then applied to these simulations to calculate its intrinsic error, basically associated with the accuracy in the calculation of the points where the BCG--to--ICL transition occurs.  The ICL maps in both filters are displayed in Figure \ref{ICLim} along with the original \rxj images. 

As mentioned previously, the ICLf can be used as an indicator of the dynamical stage of a cluster of galaxies in the redshift range 0.18$\leq$z$\leq$0.55 \citep{yoli-clash,yoli-coma}. Merging (or dynamically active) clusters show a clear signature in the ICLf measured at different wavelengths: an excess in ICLf measured in the filters that correspond to the peak emission of late-A/early-F type stars, hereafter called blue ICL excess (\textbf{BIE}). The relaxed clusters displayed a constant ICLf (within the error bars) independently of the optical band used to measure it. 
The BIE is consistent with being produced by the stripping of relatively younger and shorter lived stellar populations in the outskirts of galaxies into the ICL during merger. In this scenario, it would be expected that the A--F brighter population responsible for the BIE would leave the main sequence towards the giant locus effectively vanishing from the ICLf in the frequencies corresponding to the BIE, within a relatively short timescale after being stripped. Roughly, given a half-life for this stellar population of about $\sim$ 2 Gyr (for F2 star) after being stripped it would be reasonable to assume that that would be the duration of the BIE, after which, we would observe the overall ICLf wavelength distribution similar to that of 
relaxed clusters, i.e., a flat distribution profile (Figure \ref{all_clusters}). 

In Figure \ref{all_clusters}, in addition to all ICLHST clusters, we also included a recently measured merging system, WHL J013719.8-08284 at z=0.566 \citep{yoli-whl}, two Frontier Fields clusters with ICL analyzed very recently, A370 (z=0.375) and S1063 (z=0.348) \citep{nicolas} and a potential “fossil system”, where a somewhat similar ICL analysis has been performed \citep{yoo}, RX J105453.3+552102 at z$\sim$0.47. 
Compared to the systems that we analyzed in the above mentioned previous works, in Figure \ref{all_clusters} it can be seen that \rxj~ behaves in a different way from all clusters: it does not show either a constant ICLf or the previously described BIE. Even though the value obtained for the F435W band is consistent with the relaxed clusters in ICLHST, the ICLf computed for the F606W band is higher than the typical values found for the passive/relaxed systems. In addition, the measured ICLf for the F606W band is significantly higher when compared to that of F435W. That particularity could, in principle, be expected if the system has been without mergers for a very long time. If the system continued to be undisturbed for a very long time one would expect that tidally stripped stars will be of earlier types, and since all the new star formation would happen inside galaxies, the ICLf in the bluer band would be reduced and in the redder band enhanced, as we observe for this system.\\

\subsection{X-Rays}
\subsubsection{Spectral analysis}

We originally extracted spectra from five annular regions within the central 600 kpc: 0--50 kpc, 50--100 kpc, 50--150 kpc, 150--300 kpc and 300--600 kpc. At the redshift of \rxj , 1$^{\prime\prime}\sim$ 2.05 kpc, so that the central bin was chosen to cover a significant region of XMM's PSF with $\gtrsim$80\% of the encircled energy fraction (see XMM-Newton Users Handbook\footnote{xmm-tools.cosmos.esa.int/external/xmm\_user\_support/documentation\\/uhb/onaxisxraypsf.html}). \cite{eric-sample} also analyzed a \rxj~Chandra snapshot observation
that we use for comparison. They found a gas temperature of $T_X=2.60^{+0.63}_{-0.53}$ keV including all emission within 250 kpc of the center. This previous Chandra observation showed that the AGN contamination is small, with $\approx$ 4\% of the central bin (0-50 kpc) and is limited to $r\leq~2^{\prime\prime}$. Spectral fittings of the central region with and without a central $8^{\prime\prime}$ did not show significant differences. The second and third bins cover the predicted cooling radius originally estimated using a $T_X \sim 2.6$ keV.
A similar modelling as that described in section \ref{sec:bkg_treatment} for the outer region was used for each of the regions, but with the best fit slopes of the SP frozen at the best fit values found in the outer region. The normalizations were free to vary. Energy bands were restricted to 0.3--8.0 keV. Overall, 24 parameters were free to vary, as opposed to the fits for the outer region designed to measure the background, where 28 parameters were allowed to vary.

We fit MOS 1,2 and PN spectra simultaneously for each observation since the results were consistent within all 3 observations individually. The best fit values of temperatures and abundances are shown in Table \ref{bestfit} and plotted in Fig. \ref{myT-A}. Both temperature and abundance profiles show negative radial gradients. The intracluster gas temperature reaches 3.0$\pm$0.1 keV in the central 50 kpc, dropping to 1.8$\pm$0.1 keV at $\ge$ 150 kpc. The metal abundance is very high in the most central bin within 30 kpc reaching 1.3$\pm0.2$ solar dropping steeply by a factor of three at 75 kpc still within the hot core. The abundance then flattens outwards at $\sim$ 0.25 solar. Central abundance gradient are fairly typical of cool core clusters, \citep[e.g.,][]{degrandi2004} but very rare in non-cool core clusters, which are typically dynamically active. We are unaware of any non-cool-core cluster having such a steep abundance gradient as that observed in \rxj.

\begin{figure}
\centering
\includegraphics[width=8cm]{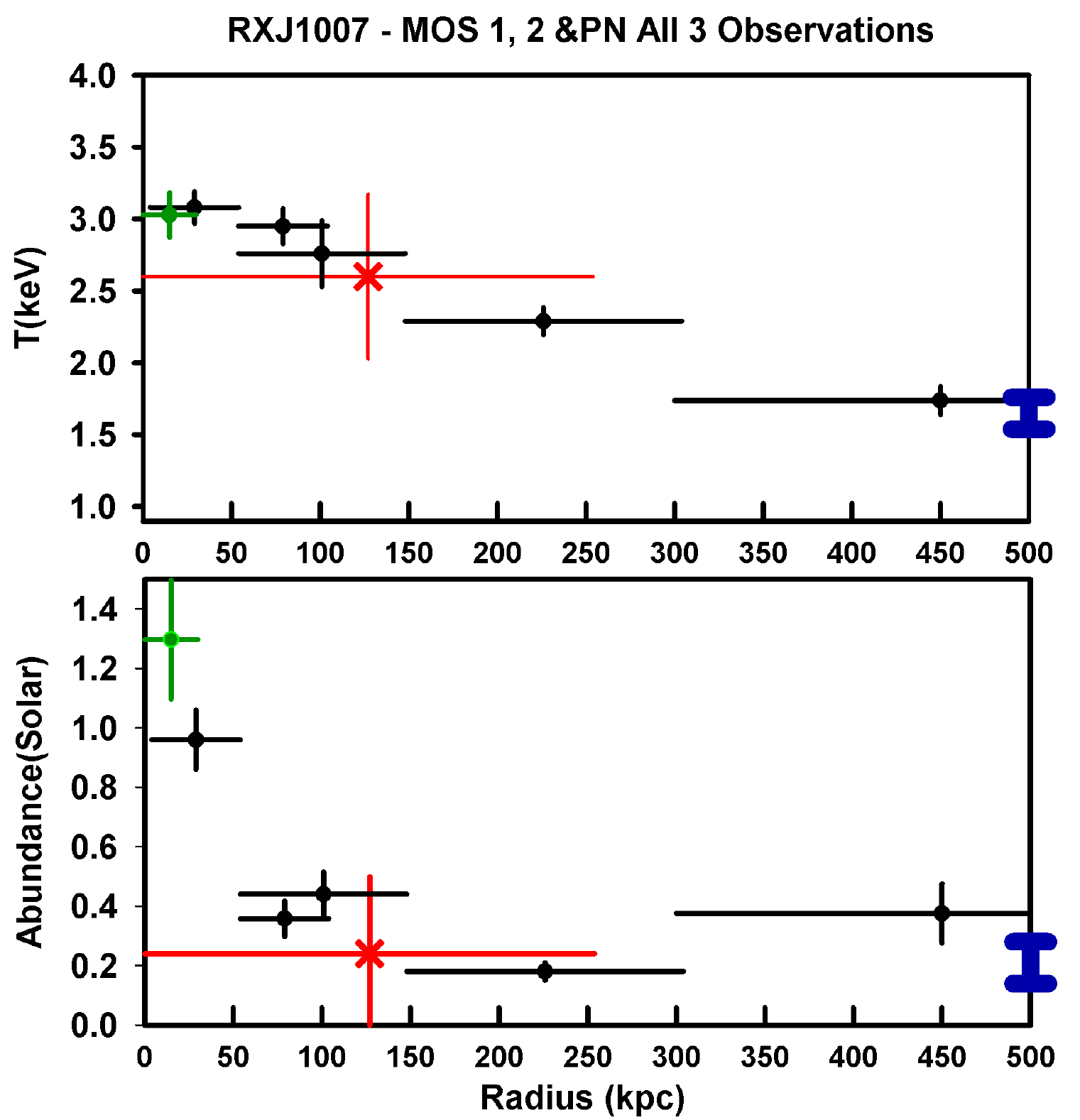}
\caption{Temperature (in keV) and metal abundance  (in solar photospheric) radial profiles for \rxj. The stars with red errors show the values derived from a previous Chandra snapshot \citep{eric-sample} for comparison. The values corresponding to the external region used to derive the soft proton contamination parameters are shown in dark blue with thick lines. The green most central bin was used to help estimate the upper limit time for the last merger in the system (see Section 4 for details.) }
\label{myT-A}
\end{figure}

\begin{table}
\centering
\begin{tabular}{ |p{1.5 cm}||p{1.5cm}|p{1.5cm}|p{2.5cm}|  }
\hline
 \multicolumn{4}{|c|}{Best-fit parameters} \\
 \hline
Bin & T$_X$ &Abundance & $\chi^2_{red} $\\
 (kpc)&(keV)&(solar)& Obs1,2,3\\
 0--30&3.08$\pm$0.15&1.33$\pm$0.21& 1.06, 1.7 ,1.2\\
0--50&3.0$\pm$0.10&0.92$\pm$0.08& 1.13, 1.13 ,0.96\\
50--100&3.03$\pm$0.10&0.43$\pm$0.06& 0.91, 1.19, 0.98\\
50--150&2.88$\pm$0.07&0.41$\pm$0.04& 1.06, 1.05, 0.95\\
150--300&2.45$\pm$0.09&0.28$\pm$0.03& 0.97, 1.16, 1.09\\
300--600&1.78$\pm$0.10&0.24$\pm$0.06& 1.00, 1.07, 1.07\\
Outer &1.72$\pm$0.08&0.4$\pm$0.12& 1.14, 1.19, 1.27\\
\hline
\end{tabular}
\label{bestfit}
  \caption{Error--weighed average of the three observations for all instruments. The reduced $\chi^2$values are shown for each region of each observation separated by commas  }
  \end{table}

\begin{figure*}
\centering
\includegraphics[width=\hsize]{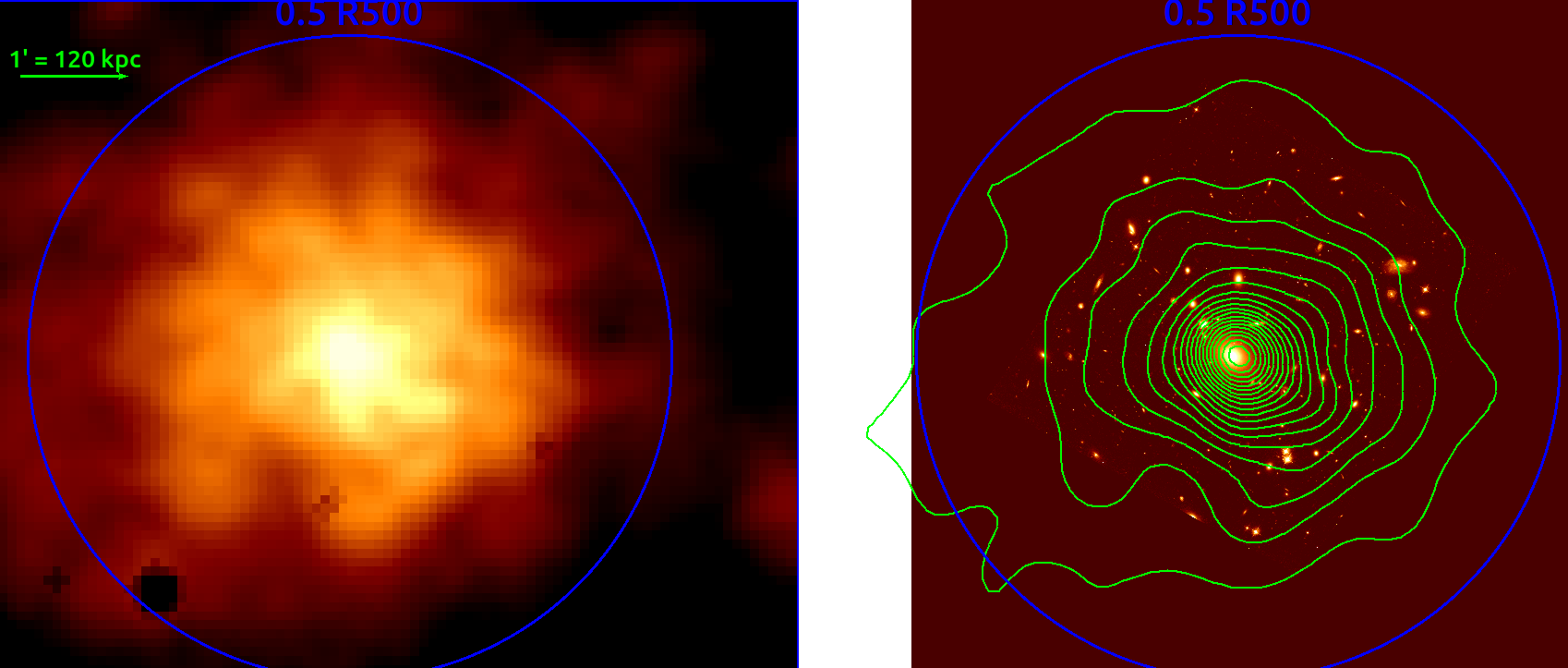}
\caption{\textbf{Left}-Full-background-subtracted and exposure-corrected image combining all EPIC detectors for \rxj. The estimated $\frac{1}{2}$~R$_{500}$ is shown in blue and the scale is shown in green. \textbf{Right} - X-ray contours overlaid on the HST image (F606W).
The 90\% encircled total energy scale based on XMM PSF is $\sim$45$\arcsec$, 47$\arcsec~$and 49$\arcsec~$at 0$\arcmin$, 1.5$\arcmin~$and 3$\arcmin~$from the center (xmm-tools.cosmos.esa.int/external/xmm\_user\_support/documentation/uhb/offaxisxraypsf.html). N is up, E is left}
\label{niceimage}
\end{figure*}

\subsubsection{Image analysis and surface brightness}
Having followed the standard procedure for background determination and determined the main characteristics (normalizations and slopes) of the SWCX and SP contamination together with the  0.4--1.25 keV QPB image we can produce a ``clean'' image for analysis. We used the tool \textit{proton} with the parameters for the broken power law and corresponding normalizations to obtain the SP images. We used the tool \textit{swcx} with the previously determined normalizations for the lines at 0.56 keV and 0.65 keV to obtain SWCX images. After aligning them to the sky coordinates using the tool \textit{rot-im-det-sky} we created a full-background-subtracted and exposure-corrected image combining all instruments using the tool \textit{comb} (with thresholdmasking parameters set to 0.02 and binning to 2). We show its smoothed version in Fig. \ref{niceimage}.

\begin{figure}
\centering
\includegraphics[width=8cm]{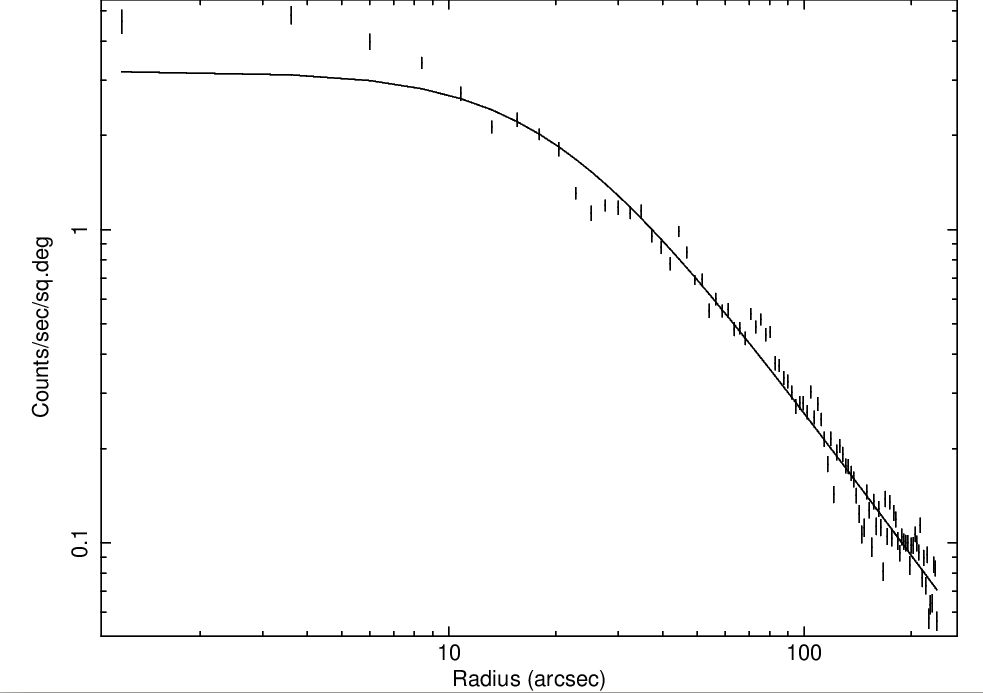}
\caption{Surface brightness profile for the full-background-subtracted and exposure-corrected X-ray image combining all detectors for \rxj~  observation 0653450201. We show here a single beta-model profile fit}
\label{100bins}
\end{figure}

We extracted the surface brightness from 100 equally-spaced annuli centered at the X-ray center. We fit these surface brightness profiles to a single beta model which takes the form of S = S0 $[1+(\frac{r}{r_c})^2]^{-3\beta+0.5}$, where S0 is the normalization, r is the projected radius, r$_c$ is the core radius, $\beta$ is the slope. The best fitting values are r$_c=39.7\pm1.5$ kpc and $\beta=0.42\pm0.003$ for a 
$\chi^2$ of 180 and a correlation coefficient of -0.69. The plot is shown in Fig. \ref{100bins}. It can be seen that the central bins present a significant excess, which is characteristic of cool core clusters. If we limit the inner fitting region such as to encompass to $>$ 50\% of the encircled energy fraction of the XMM PSF \footnote{xmm-tools.cosmos.esa.int/external/xmm\_user\_support/documentation\\/uhb/onaxisxraypsf.html} or $\gtrsim 15^{\prime\prime}$ we obtain r$_c=53.3\pm1.5$ kpc and $\beta=0.44\pm0.003$ for a 
$\chi^2$ of 175 and a correlation coefficient of -0.77. This is fully consistent with the values obtained by the Chandra snapshot by \citep{eric-sample}, show found $r_c=50^{+19}_{-15}$ kpc\footnote{with a 0.44\% offset if adjusted for the assumed cosmology}, $\beta=0.5^{+0.09}_{-0.07}$ in a single $\beta$-model surface brightness fitting. 

\subsubsection{Mass estimates}
The masses can be derived using Euler's momentum equation 

\begin{equation} \label{euler}
\rho \frac{\partial \textbf{v}}{\partial t} + \rho (\textbf{v}.\nabla)\textbf{v} = -\nabla P+ \rho\textbf{g}
\end{equation}
which, in the absence of bulk fluid velocities (\textbf{v}) 
allows the gravitational field (\textbf{g}) to be fully characterized by the gas density ($\rho$) and pressure (\textbf{P}) profiles. As implied from the single beta models used to fit the surface brightness, the particle density profile as a function of the core radius, central density ($n_0, r_c$), is expressed as  $n=n_0~(1+(\frac{r}{r_c})^2)^{-3\beta/2}$. Following \cite{arnaud99} we call this the $\beta$ model (BM) approach. For an isothermal case, the mass enclosed within a particular radius can be expressed as 

\begin{equation} 
\label{mbm}
M(\le r)=1.11\times 10^{14}\beta~ (\frac{\mu}{0.6})^{-1} \frac{kT}{keV} \frac{r}{Mpc} (\frac{(\frac{r}{r_c})^2}{1 + (\frac{r}{r_c})^2}) M_{\odot}
\end{equation}
where $\mu$ is the mean molecular weight\footnote{we leave $\mu$ explicit because it is likely that its value is significantly higher than 0.6 throughout the regions usually probed in X-rays \citep{pessah2013}}.
and

\begin{equation} 
\label{rdelta}
r_{\delta_c}=0.627\beta^{\frac{1}{2}}(\frac{\mu}{0.6})^{-\frac{1}{2}} (\frac{\delta_c}{500})^{-\frac{1}{2}}(\frac{kT}{keV})^\frac{1}{2} Mpc
\end{equation}

where $\delta_c$ is the density contrast  with respect to the critical density.
Assuming that the hot core-excised temperature is the value at and over 300 kpc, i.e. 1.78 keV, and the best fit values for r$_c$ and $\beta$ from surface brightness fitting (previous section) we obtain M$_{500_{BM}}=(0.39\pm0.05)\times 10^{14}$M$_{\odot}$, where r$_{500_{BM}}=0.53\pm0.02~$Mpc.
To correct for the central negative temperature gradient, we can assume that the temperature distribution can be approximated by two isothermals, the first one with the internal temperature set at the value correspondent to the error weighted average of the regions at $\lesssim$ 200 kpc, i.e., the hot core, (2.78$\pm$0.05) keV, and a second isothermal with our most external region value of (1.78$\pm$0.1) keV. We then add the mass difference  from the two isothermals measured in the inner hot core to the total mass derived with the second isothermal and find M$_{500}=(0.47\pm0.06)\times 10^{14}M_{\odot}$ \footnote{A single isothermal with T$_X$=2.78$\pm$0.05 would give M$_{500}=(0.61\pm0.08)\times 10^{14}M_{\odot}$}

A more straightforward way to estimate the mass throughout the non-isothermal region would be to directly fit P(r) within the ``hot'' region ($r\le r_{hot}$), so that 
\begin{equation} 
\label{pressure}
M(\le r_{hot})=\frac{-dP/dr}{G\mu m_H n_{hot} r_{hot}}
\end{equation}

where $n_{hot}$ is the number density at r$_{hot}$, G and m$_H$ are the gravitational constant and the H mass.
Using a linear fit for P(r), we obtain  M$_{500}=(0.52\pm0.06)\times 10^{14}M_{\odot}$

On the other hand, one can also estimate the system mass using scaling relations based on virialization (VT) either through X-ray temperature \citep{gus1996} or galaxy velocity dispersion measurements \citep{carlberg1997} as in equation \ref{eq1}. For the former, adopting $r_{500}=0.56(\frac{kT}{keV})^\frac{1}{2} \frac{H(z)}{H0} Mpc = 0.75\pm 0.06$ Mpc, which is very close to that derived from the Chandra snaphot \cite{eric-sample}, we get M$_{500}=(0.64\pm0.08)\times 10^{14}M_{\odot}$ (M$_{200}=(0.95\pm0.11)\times 10^{14}M_{\odot}$)\footnote{using r$_{200}$=1.54 r$_{500}$\label{r200500}}.
We chose to use this value for the mass henceforth throughout the paper because (1) it is the most consistent with that derived from velocity dispersion (section \ref{sec:gemdata}), (2) it is less prone to systematics involved in the estimation of the surface brightness parameters with XMM and (3) it provides the most conservative value in the interpretation of the specific ICLf peak at 5400\AA, which will become clear in the next section.

Integrating the density up to r$_{500}$ we obtain a gas mass of $M_{gas_{500}} \sim (7.59\pm0.24)\times10^{12}M_{\odot}$ for $\mu$ = 0.6.
This would imply a gas fraction f$_{gas_{500}}=0.119\pm0.041$  (f$_{gas_{200}}=0.169\pm0.06$)\textsuperscript{\ref{r200500}}.
Assuming a galaxy ($M/L_r$) ratio of 6 in the r-band, obtained converting from the i-band in \cite{cappellari2006} for a $(r - i)$ = 0.5, we obtain the mass from galaxies M$_{gal_{500}}= (4.36\pm0.02)\times 10^{12} M_{\odot}$ and from the ICL M$_{icl}\sim (0.47\pm0.003)\times 10^{12} M_{\odot}$, where the latter is a lower limit, since we conservatively estimated only up to the radius that the ICL could be measured, that is, for r$\leq$ 178 kpc and did not extrapolate the ICL beyond that. This would bring the total baryon fraction\footnote{This estimation does not include lesser order such as HI or H2 masses \citep[e.g.,][]{morganti2006}}  f$_{b_{500}}$ to 0.19$\pm$0.03. This is higher than all groups of similar masses analyzed in \cite{tatiana} and more compatible with higher mass clusters ($\sim 1.5-4)\times 10^{14}M_{\odot}$. The ratio of stellar mass to total mass within r$_{500}$,  f$_{*500}=\frac{M_*}{M_{500}}$, when the ICL is included is found to be $\sim$ 0.075$\pm$0.008, again very high and compatible with much more massive clusters \citep{tatiana}. The mass to light ratio within r$_{500}$ is consequently low and is found to be ($\frac{M_{500}}{L_r})=82\pm10(\frac{M}{L_r})_\odot$, lower by a factor of $\sim$2 than the more massive REXCESS clusters, but more consistent their "REGULAR" systems \citep{rexcess15}. If the mass-to-light ratio does not vary significantly from r$_{500}$ to r$_{200}$ the system would not stand out as a "dark" system, as suggested by the FG sample studied by \cite{rob}

The entropy (S$_X$) calculated as S(r) = kT(r) n$_e(r)^{-2/3}$ at 0.1~r$_{200}$ is found to be 204$\pm$71 keV cm$^{2}$. Assuming the error-weighted average of all independent regions (2.59$\pm$0.04 keV)\footnote{This is extremely close to that measured by \cite{eric-sample} } as representative for this system, the value is consistent with that found for other FGs , lower than that for non-FG groups, and it is closer to the extrapolation of self-similarity defined by massive clusters \citep{khosroshahi2007}. In the most central region the entropy is S($\sim0.02r_{200})=96\pm34$ keV cm$^{2}$, lower and with a shallower profile than that of the hot FG RX J1416.4+2315 \cite{habib06}, but consistent with the expected entropy floor for its overall temperature \citep{ed2000} and marginally more consistent with the class of systems with entropy floor $\leq$ 50~keV cm$^{2}$ as defined by \cite{cavagnolo}.\\

\section{Discussion}
The standard model of FGs, as being relaxed and undisturbed systems, necessarily leads to the production of well developed cool cores. In this work we confirmed that \rxj~ does not have a cool core. The radial intracluster gas temperature gradient is actually negative, the opposite of what one would expect in a cool core system. On the other hand, the ICM shows a steep negative radial abundance gradient typical (perhaps exclusive) of cool core clusters and groups,
where the abundance may reach super solar values in the center \citep{ettori}. Furthermore, its surface brightness also hints a departure from a 
single $\beta$ model fitting in the central 15\arcsec, again, typical of cool core clusters.

Another cool core characteristic of \rxj~ can be inferred by checking concentration parameters, in particular the so-called ``surface brightness concentration''. We denote it here as c$_{ISB}$, since it is in reality an Integrated Surface Brightness (or flux) over some ``core'' region divided by that over some larger region. \cite{santos2008} proposed this parameter when analyzing 26 clusters at low and high redshifts. They found that cool-core clusters had c$_{ISB}\geq0.075$ using  (inner/outer) radii of (40 kpc/400 kpc), respectively. Using the same radii for \rxj~ by direct background subtracted counts in these regions we obtain c$_{ISB}=0.1\pm0.03$. Even if we used the best-fit $\beta$ model fit, which tends to underestimate the  central flux, we obtain c$_{ISB}\geq~0.078\pm0.013$, still consistent with the limit for cool core systems. 
In any case, this suggests that the core gas concentration is not as disrupted by the last merging event as the core temperatures.

The strongest indication of long age for \rxj~ comes from the ICLf analysis (Figure \ref{all_clusters}), which shows a color distribution that is not consistent with merging (dynamically active) clusters but compatible with that observed in ``relaxed clusters''. It shows indications of being even older than the relaxed clusters analyzed so far, based on the high ICLf found in the F606W filter. 
As mentioned in section 1, the specific ICLf, or \(\frac{ICLf}{Mass}\) ratio, hereafter denoted by \textbf{ICLf$_M$}, may be an indicator of relative age of the system. With a mass of $\sim 0.64\times 10^{14} M_{\odot}$ within r$_{500}$ we measure significantly higher ICLf$_{M_{500}}$ ratios in comparison to all clusters, both relaxed and merging,
especially in the redder filters. We found for \rxj~ICLf$^{F435W}_{M_{500}}$=~(114$\pm$56) $(10^{15}M_{\odot})^{-1}$ and  ICLf$^{F606W}_{M_{500}}$=~(193$\pm$24) $(10^{15}M_{\odot})^{-1}$ for the F435W and F606W filters, respectively, as can be seen in  Figure \ref{rest-IM}, where we plot the rest-frame ICLf$_{M_{500}}$ for all systems showed in Figure \ref{rest-IM}.
It can be seen in that figure that \rxj~ICLf$_{M_{500}}$ stands out from \textit{all} clusters, especially at its reddest measured band at $\lambda\sim$5400 \AA, where it is more than five times higher than perhaps the most relaxed cluster in the ICLHST sample and closest in redshift and mass, A383 (z=0.187), which in turn has an ICLf$_{M_{500}}$ more than twice as high as those measured for the intermediate-z merging clusters MACS J0717.5+3745 (z=0.548), MACS J1149.5+2223 (z=0.544) and WHL J013719.8-08284 (z=0.566) at that band. 

 \begin{figure*}
\centering
\includegraphics[width=\hsize]{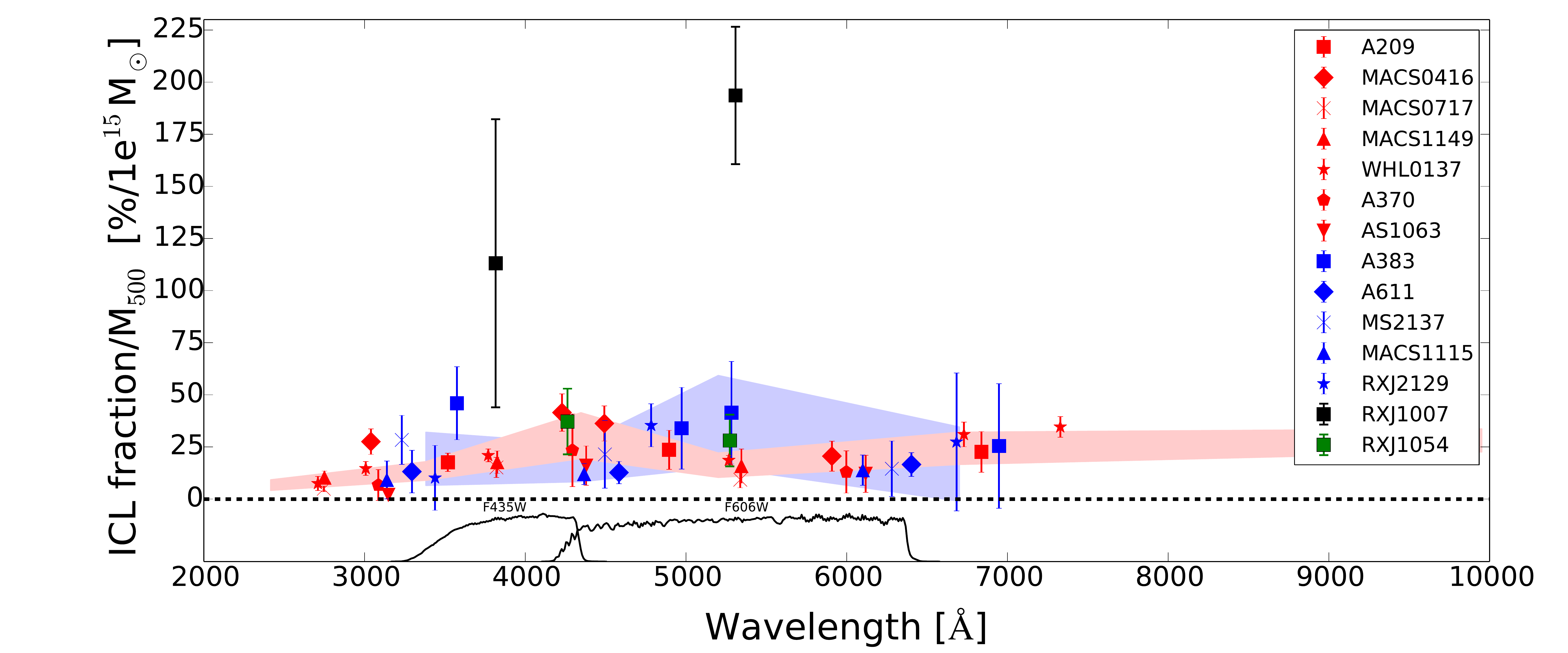}
\caption{Rest-frame ICLf/M$_{500}$ ratio comparison. In black we show the rest-frame ICLf values derived in this work for \rxj~ (black), for all relaxed (blue) and merging (red) ICLHST clusters analyzed in \cite{yoli-clash} as well as for the Frontier Field clusters AS1063 and A370 \citep{nicolas}, the high-z merging cluster WHL J013719.8-08284 \citep{yoli-whl} and a candidate fossil cluster RX J105452.03+552112.5 in green \citep{yoo}. We show the transmission curves of the HST filters used to determine the ICL in \rxj~ in the bottom. The error shaded bands are the same as in Figure \ref{all_clusters}.}
\label{rest-IM}
\end{figure*}

The unusually enhanced ICLf$_M$ in the longer wavelengths suggests that the FG has been injecting ICL through the steady regime without increasing its mass for a long period of time. This is what would be expected if the system reached the end of its merging tree and has been dynamically undisturbed since then. We can estimate the maximum time of the last merger from cooling time constraints. The central cooling time can be estimated from Equations 7 \citep{voigt2004}. For that estimation, we used a smaller (r=15\arcsec) central region, chosen as the minimum region where we could still have enough counts to determine T$_X$ with reasonable precision, given the relatively large XMM PSF. The results are also shown in Table \ref{bestfit} and plotted in Figure \ref{myT-A}. 
The temperature there is found to be T$_X\sim$ 3 keV}. 
In that region, we derive a central density of $(6.7\pm0.9)\times 10^{-3} cm^{-3}$, corresponding to t$_{cool}=(4.8\pm0.66)$ Gyr, if we use the average value of the equations 7 for temperatures over and under 3 keV. This value also satisfies one of the criteria for a cool core system \cite[e.g.][]{hudson2010,birzan2004,su2016}.
\begin{equation}
\begin{aligned}
t_{cool} \sim
 \begin{cases}
  20(\frac{n_e}{10^{-3} cm^{-3}}) (\frac{T_X}{10^7 K)})^{\frac{1}{2}} Gyr & for~  T_X\geq3 keV\\
  5 (\frac{n_e}{10^{-3} cm^{-3}}) (\frac{T_X}{10^7 K})^{\frac{8}{5}} Gyr & for~  T_X\leq3 keV\\
\end{cases}
\end{aligned}
\end{equation}

The lack of a cool core in that very central region also sets the cooling time as the upper limit for the last merger event in this system. It should be noted that this value is more than twice the time needed to the late-A early-F stars injected in the ICL during a merger to evolve away and erase the BIE feature, given the average half-life of $\sim$ 2 Gyr for a main-sequence F2 star. In principle, the BIE lifespan could gives a rough time scale corresponding to the lower limit for the last merging event. One caveat with this lower limit is that we are assuming that there would be a BIE resulting from fast stripping of the "younger" stars from the outskirts of galaxies, particularly late-types in any merger. Certainly the situation is likely to be much more complex, perhaps involving full tidal disruption of galaxies. We did include one ``fossil cluster'', the only one where a similar ICL analysis has been performed, using Gemini data in i and r bands, RX J105453.3+552102 at z$\sim$ 0.47, in Figures \ref{all_clusters} and \ref{rest-IM}  \citep{yoo}. It has a dynamically estimated mass of M$_{200}\sim 8\times 10^{14} M_{\odot}$ and it presents a somewhat high ICLf at the BIE peak, but has a flat profile within the (large) errors. This system is, however, unusual, showing signiﬁcant departure from Gaussianity of group member velocities \citep{aguerri} and a large mismatch of the X-ray centroid and the BCG, which would be consistent with a recent merger.

The ICLf$_M$ of \rxj~peaks roughly (given the F606W width) at $\sim$ 5400 \AA~ (4300\AA-6500\AA) encompassing the peak emission of G-main sequence stars. Once the late type stars in the member galaxies outskirts are removed, it is plausible that tidal stripping would start to removing earlier type populations closer to the central regions of the galaxies, which differently from those stars responsible for the BIE, would be redder and have significantly longer lifespans, assuming that processes of galaxy rejuvenation \citep[e.g.,][]{fang,mancini2019,paola} are not significant.

This ICLf$_M$ enhancement allows us to have a rough idea of the lower limit for an ICL injection rate during the steady regime for ICL generation. To be conservative, we compare here the ICLf$_{M_{500}}$ of \rxj~ to that of A383 at the  ICLf$_M$ peak wavelength, since it has the closest ICLf$_M$ value to that of \rxj~ at that wavelength. If we denote
\begin{equation} 
\label{eq4}
(\frac{ICLf}{M_{500}})_{RXJ1007} = \eta~ (\frac{ICLf}{M_{500}})_{A383}
\end{equation}

and assume that $\eta\sim 1$ at the time of the last merger event in \rxj~ (t$_{merge}$) and that the ICL continued to be injected in a passive steady regime up to its current redshift, without increasing the system's mass, its injected ICL rate is

\begin{equation} 
\label{eq5}
\frac{\Delta L_{ICL}}{\Delta t}=\frac{L_{ICL_{obs}}(1-\frac{1}{\eta_{obs}})}{\Delta t}
\end{equation}

where the subindex \textit{obs} indicates the observation epoch, t$_{merge}+\Delta t$ yr, and L stands for luminosity. We also assume that the ICL injection and the galaxy luminosity loss happen in the same proportion, i.e., $\Delta L_{ICL}=-\Delta L_{gal}$, which ignores significant star formation during that time. However, the discrepancy with this assumption can be alleviated keeping in mind the outside-in quenching in cluster galaxies, backed up by observations \citep[e.g.,][]{johnston2014} and numerical simulations \citep{pfeffer2022}.  
Using the observed (Fig \ref{rest-IM}) $\eta_0$ of 5.8$\pm$2.4 and the same (M/L) ratio as in the previous section  we obtain a $(\frac{dM_{ICL}}{dt})_{r\leq178kpc}$ of (98$\pm$43) M$_\odot$ yr$^{-1}$,
for t$_{merge}$ = t$_{cool}$. Again, conservatively, we carry out this estimation only up to $\leq$ 178 kpc, the radius at which the ICL could be measured confidently.

\cite{tollet2017} has predicted loss of stellar mass of satellite galaxies entering groups and clusters due to tidal stripping using cosmological N-body simulation and abundance matching technique \citep[][]{AM2007} for two different modes: \textit{shutdown} and \textit{starvation}. In the former, the gas available to form stars is fully removed upon entry into the host halo. In the latter, star formation continues after entry. In the case of massive systems (M$_{500}\ge10^{14}$M$_\odot$) from \cite{gonzalez2013}, their work favored the starvation plus tidal stripping mode. In \rxj~ the ratio of the BCG+ICL masses to the total stellar mass (M$_{BCG}$+M$_{ICL}$)/M$_*$ within r$_{500}$  is found to be 0.53$\pm$0.003, which significantly disagrees with the prediction for the shutdown model and, instead, is in very good agreement with the starvation + tidal stripping model found for the more massive systems (Figure 14 of \cite{tollet2017}). Specific simulations of these combined processes, i.e., changes in radial age gradients in cluster galaxies and ICL injection will be very helpful to assess the differences in the ICL flux distribution in FGs.

The very high intracluster gas central abundance enhancement in \rxj~is also puzzling. It would likely require a pre-existent high abundance gradient in at least one of the pre-merging systems. A situation similar to the case of Abell 1142 \citep{su2016}, where the merging BCGs would subsequently settle in the high metal core. Even though metal transport through BCG sloshing have been observed \citep[e.g.,][]{dupke2007, simionescu2010}, it is not physically implausible that central abundance gradients could survive mergers, 
 even if the cool cores were destroyed, given the different nature of the mechanisms of thermalization and chemical mixing and the inefficiency with which post merging sloshing mixes the gas \cite{simona14}.

Numerical simulations of cluster mergers can provide a few helpful hints with respect to the nature of the merger. To keep the central temperature high for such a long time and avoid a metal abundance mixing, small impact parameter mergers seem more likely. For example, \cite{ricker2001} work shows a core \textit{reheating} prior to the core merging, about 4 Gyr after the first core-crossing, for an impact parameter b$\lesssim$ 2 r$_s$, where r$_s\sim$ 140 kpc\footnote{for what seems to be the closest merging configuration to ours see ``\textit{AA}'' in \cite{ricker2001}} is the Dark Matter scaling radius. 

This time scale is also consistent with the merging of the BCGs. \cite{kitz2008} use a semi-analytic model to estimate the merging rate and timescale of galaxy pairs, based on the Millennium N-body simulation. They ﬁnd that the merging timescale is T $\sim$ 3.2 r$_p$ $M^{-0.3}_*$ Gyr for galaxies, where the line of sight velocity difference between the two galaxies is less than 3000 km s$^{-1}$, where r$_p$ is the maximum projected separation of the two galaxies in units of 50 h$^{-1}_{70}$ kpc and M$_*$ is the stellar mass of the galaxies in units of 4$\times$10$^{10}$h$_{70}^{-1}M_{\odot}$.
Using the luminosity of \rxj~  BCG and a $\frac{M}{L_i}$ ratio in the i-band of 3.8 \citep{cappellari2006} we get a BCG merging time of $\sim$ 3.4 Gyr for an r$_p \leq 2r_s$,  assuming that each pre-merging BCG would share the mass of the post-merging BCG equally.

Independent of the selection uncertainties using the magnitude gap criterion, there is some consensus that systems that assembled a significant fraction of their final mass earlier ("old") will tend to develop large magnitude gaps. Its been argued frequently that the gap is unlikely (10\%) to survive for longer periods ($>$ 4Gyr) of time \citep[e.g.,][]{benda2008,dariush}, even though this fraction can be substantially higher for systems over 10$^{14} M_{\odot}$ \citep[][]{goza14b}. This work shows that \rxj~ has survive as an FG for $\sim$ 5Gyr, and it will survive significantly longer given its wide $\Delta m_{1,2}\sim$ 2.5 (Figure \ref{spatd}). So, the question of how a FG would be able to maintain itself as an FG (under the magnitude gap $\Delta_{m_{1,2}}$ criterion) over its evolutionary history is pertinent. Given the typical presence of several bright ($\Delta_{m_{1,2}}\leq2$) galaxies within 0.5 r$_{200}$ in galaxy clusters, it is easy to imagine that the merger of an FG with a galaxy cluster would result in a cluster and not in a FG. One could in general expect that FG+Cluster=Cluster and Cluster+Cluster=Cluster. 

One interesting possibility for long term survival of FGs is through FG-FG merger. Something shown in detail only once so far for the Cheshire Cat \citep{jimmy-cat}. The redshift difference between the two BCGs, which represent the “eyes” of the Cat, corresponds to 1350 km s$^{-1}$ in the group's rest-frame. In a combined Chandra-HST-Gemini analysis of the system, \cite{jimmy-cat} showed compelling evidence that that system is actually the merging of two separate galaxy groups, both qualifying (by the  $\Delta_{m1,2}$ criterion) as FGs\footnote{One of the merging FGs ``G2'' in \citep{jimmy-cat} could be considered as an FG candidate rather than an FG since the magnitude difference between the first and second rank galaxies of that merging subsystem is 1.83, nearly missing the ``default'' $\Delta_{m1,2}$=2.0.}, which are in collision course and will merge in $\sim$0.9 Gyr, becoming again a larger FG. Analogous to \rxj, that system also has a negative temperature gradient going from $\sim$ 5.4 keV in the central 115 kpc to $\sim$3 keV in the outer parts. The high X-ray temperatures measured near the center of the Cheshire Cat are the result of shock heating from the merger of two FGs. 

The existence of the Cheshire Cat highlights the potential importance of fossil group progenitors to explain the permanence of FGs over time even in the presence of merging. A search for these so called fossil progenitors in the CASSOWARY strong gravitational arcs catalog found $\sim$13\% of the lensing groups were identiﬁed as FGs and that $\sim$23\% of lensing systems were fossil progenitors, 6\% higher than in the control sample. The CASSOWARY systems are a good place to look for them because strong gravitational lensing preferentially selects systems with a high mass concentration such as fossil systems \citep{lucas-fpchandra}. 

In fact, many CASSOWARY fossil progenitors show highly asymmetric BCGs along with higher X-ray luminosities and ICM temperatures than group scaling relations predict \citep{lucas-fplense} implying a higher than expected gravitational potential based on group richness.
Fossil progenitors appear to be in mid-assembly of their dominant BCG (akin to the Cheshire Cat) making these excellent candidates to follow up observations of their ICLf to check for the presence of BIE and ICLf$_M$. This would allow us to see the evolution of these properties prior to a ``final'' merger and to compare them to that of FGs.

Near future spectrophotometric surveys with high precision (0.3\%) photo-z such as J-PAS, which has 54 narrows-band contiguous filters \citep{jpas,minijpas}, will allow us to measure the ICL SED for a very large number of systems up to relatively high redshifts z$\lesssim$0.6--0.8, which will provide invaluable information to the age and dynamical state of FGs.

\section{Summary}
We have analyzed X-ray, optical (HST) and spectroscopic (Gemini) data of a classic FG \rxjfull, which presents many of the contradictory physical properties often found in these systems. We performed a combined multi-filter ICLf analysis in a bona fide FG to test its dynamical state and age and compared to other systems merging and relaxed, where similar analyses have been carried out. We found that:   

\begin{itemize}
\setlength\itemsep{0.7mm}
\item  the absence of a cool core in \rxj~was corroborated. We instead found a hot core, where the intragroup gas temperature rises to $\sim$ 3 keV in the central 30 kpc, and drops to $\sim$ 1.8 keV for r$\geq$300 kpc. The central metal abundance is very high, reaching supersolar values (1.3 solar) in the central 30 kpc and dropping very steeply  at 75 kpc still within the hot core. 

\item the system's mass derived from hydrostatic equilibrium assumption is found to be  M$_{200X}=(0.95\pm0.11)\times 10^{14}M_{\odot}$ (M$_{500X}=(0.64\pm0.08)\times 10^{14}M_{\odot}$), which is consistent with the mass derived from galaxy velocity dispersion M$_{200g}=(1.30\pm0.35)\times 10^{14}M_{\odot}$.

\item the gas fraction and baryon fraction of \rxj~ are found to be f$_{gas_{500}}$ = 0.12$\pm$0.04 and  f$_{b_{500}}$ = 0.19$\pm$0.03. This value is quite high and, in general, more consistent with higher mass systems, with M$_{500}\sim (1.5-4)\times 10^{14}M_{\odot}$.

\item The stellar mass fraction,including the ICL is found to be $\sim$ 0.075$\pm$0.008. It is compatible with much more massive clusters. The mass to light ratio $\frac{M_{500}}{L_r})=82\pm10(\frac{M}{L_r})_\odot$, lower by a factor of $\sim$2 than the more massive REXCESS clusters.

\item The X-ray surface brightness concentration, using  (inner/outer) radii of (40 kpc/400 kpc), is  c$_{ISB}=0.1\pm0.03$. This is consistent with the values found  for cool core clusters (c$_{ISB}\geq0.075$).

\item the gas entropy at 0.1r$_{200}$ is found to be 204$\pm$71 keV cm$^{2}$, consistent with that found for other FGs , lower than that of non-FG groups and closer to the extrapolation of self-similarity deﬁned by massive clusters. In the most central region the entropy is $\sim96\pm$34 keV cm$^{2}$, consistent with the expected entropy floor for its overall temperature.

\item the ICL fraction wavelength distribution analysis of the FG shows the absence of the blue ICL excess (BIE) in disagreement with what was found for merging (dynamically active) clusters and closer to that of relaxed clusters. In fact, when comparing the ICLf distribution of \rxj~ to other relaxed clusters analyzed previously, one can see that the unique combination of a reduction in ``blue'' ICLf with the enhancement in ``green/red'' ICLf suggests that the system has been relaxed for a very long time.  

\item \rxj~ very old age is particularly visible when looking at the distribution of specific ICLf, or $\frac{ICLf}{Mass}$ ratio. We find for \rxj~ICLf$^{F435W}_{M_{500}}$=~(114$\pm$56) $(10^{15}M_{\odot})^{-1}$ and  ICLf$^{F606W}_{M_{500}}$=~(193$\pm$24) $(10^{15}M_{\odot})^{-1}$ for the F435W and F606W filters. This is significantly higher in comparison to all clusters, both relaxed and merging measured so far,
especially in the redder filters. This is what would be expected from a system that has not had any merger for a very long time and is producing ICL only in the steady regime. 

\item if we assume that the specific ICLf of \rxj~ was the same as that of the other clusters, prior to the last merger, an ICL injection equivalent to 98$\pm$43 M$_\odot$ yr$^{-1}$ within the central 178 kpc, would be needed to raise the specific ICLf to its current values.

\item the cooling time of the system estimated from the central (r$\leq$ 30 kpc) electron density is 4.8$\pm$0.66 Gyr. This value can be considered as an upper limit for the last merging event at z$\sim$ 0.45 the lower limit of which can be roughly estimated by the complete absence of a BIE in the ICLf as $\sim$ 2 Gyr, or about a half-life of a F0-F4 star.

\item despite the absence of a cool core, \rxj~ has characteristics of cool core systems including a high X-ray flux concentration and, in particular, a steep radial abundance gradient achieving supersolar values in the central 30~kpc and dropping outwards to 0.2 solar at r$\geq$150 kpc. This indicates that at least one of the pre-merging systems that formed \rxj~ had a central abundance enhancement, possibly both.

\item the ratio of the BCG+ICL masses to the total stellar mass within r$_{500}$  is found to be 0.53$\pm$0.003, which significantly disagrees with the prediction for the shutdown model for galaxy evolution in clusters. Instead, the results strongly favor the starvation + tidal stripping model found for the more massive systems.

\end{itemize}

Overall, this work puts forward a potential age indicator for galaxy systems using the ICL fraction, that, in this case, provided further evidence towards a longer age for FGs, even for those without cool cores. Assuming that the FG achieved the "end" of its merging tree and injected ICL only through internal dynamical friction and tidal stripping, without increasing its mass through mergers, one would expect the specific ICLf to be higher than that clusters of similar mass, but still undergoing a history of mergers and this is consistent with what we observed. Furthermore, the fact that the magnitude gap of this system has been lasting (and will last) for significantly longer times than what simulations predict justifies the search for alternative mechanisms for magnitude gap longevity. Considering that mergers in between FGs seem more likely to form an FG than cluster-cluster or FG-cluster mergers, it is possible that regions with an enhanced preference for FG formation early on would increase the chances for FG survival to z$\sim$0. Further multi-wavelength analysis of bona fide FGs and FG progenitors will shed light on the evolution of this systems.

\section*{Acknowledgements}

R.A.D. acknowledges partial support from NASA grants 80NSSC20P0540 and 80NSSC20P0597 and the CNPq grant 308105/2018-4. R.A.D. also thanks Dr. MARc Kessler for very insightful discussions, Drs. Francois Mernier and Zack Li for helpful suggestions. Y. J-T has received funding from the European Union’s Horizon 2020 research and innovation programme under the Marie Skłodowska-Curie grant agreement No 898633. Y. J-T. also acknowledges financial support from the State Agency for Research of the Spanish MCIU through the ‘‘Center of Excellence Severo Ochoa’’ award to the Instituto de Astrofísica de Andalucía (SEV-2017-0709). This study was financed in part by the Coordena\c{c}\~ao de Aperfei\c{c}oamento de Pessoal de N\'ivel Superior - Brasil (CAPES) - Finance Code 001. Based on observations obtained at the international Gemini Observatory, a program of NSF’s NOIRLab, which is managed by the Association of Universities for Research in Astronomy (AURA) under a cooperative agreement with the National Science Foundation. on behalf of the Gemini Observatory partnership: the National Science Foundation (United States), National Research Council (Canada), Agencia Nacional de Investigaci\'{o}n y Desarrollo (Chile), Ministerio de Ciencia, Tecnolog\'{i}a e Innovaci\'{o}n (Argentina), Minist\'{e}rio da Ci\^{e}ncia, Tecnologia, Inova\c{c}\~{o}es e Comunica\c{c}\~{o}es (Brazil), and Korea Astronomy and Space Science Institute (Republic of Korea). This work was enabled by observations made from the Gemini North telescope, located within the Maunakea Science Reserve and adjacent to the summit of Maunakea. We are grateful for the privilege of observing the Universe from a place that is unique in both its astronomical quality and its cultural significance.

\appendix

\section{EPIC Background Modeling with External Pointing Constraints} \label{appx:bkg_modeling}

\begin{table}[h!]
\caption{Best fit parameters for X-ray background components.}
\centering
\begin{tabular}{ccc}
	\hline \hline
	Component & $\Gamma$ or kT${\rm ^a}$ & Norm \\
	\hline
	CXB${\rm ^b}$ & 1.46 & 6.8$_{-0.1}^{+0.2}$ \\
	MW${\rm ^c}$ & 0.26 & 1.3$_{-0.2}^{+0.1}$ \\
	LHB${\rm ^c}$ & 0.10 & 9.8 $\pm$ 0.2 \\
	\hline \\
\multicolumn{3}{p{7cm}}{${\rm ^a}$ Power law photon index ($\Gamma$) for CXB component or plasma temperature (kT) component given in units of keV, for MW and LB components. These values were fixed during the spectra fitting. } \\
\multicolumn{3}{p{7cm}}{${\rm ^b}$ Normalization of the power law component is given in units of $\rm{10^{-7}\,photons\,keV^{-1}\,cm^{-2}\,s^{-1}\,arcmin^{-2}}$ at 1\,keV.}\\
\multicolumn{3}{p{7cm}}{${\rm ^b}$ Normalizations of {\tt apec} thermal component are given in units of $\rm{10^{-21}\,cm^{-5}\,arcmin^{-2}}$.}
\end{tabular}
\label{tab:bkg_params}
\end{table}

\begin{table}
\caption{Error--weighed average of the three observations for all instruments using the methodology presented in the Appendix. The reduced $\chi^2$ values are shown for each region of each observation separated by comma.}
\centering
\begin{tabular}{lccc}
\hline \hline
Bin & T$_X$ &Abundance & $\chi^2_{red} $\\
 (kpc)&(keV)&(solar)& Obs1,2,3\\
 \hline
0--51.3        &   2.95 $\pm$ 0.11 &    0.76 $\pm$ 0.10 & 1.27, 1.11, 1.85 \\
51.3--102.6    &   3.05 $\pm$ 0.11 &    0.43 $\pm$ 0.06 & 1.02, 0.89, 1.12 \\
51.3--147.7  &   2.92 $\pm$ 0.08 &    0.40 $\pm$ 0.04 & 1.08, 1.04, 1.11 \\
147.7--307.8 &   2.69 $\pm$ 0.10 &    0.18 $\pm$ 0.03 & 1.07, 1.20, 1.05 \\
307.8--615.6   &   1.63 $\pm$ 0.07 &    0.09 $\pm$ 0.01 & 1.08, 1.39, 1.18 \\
\hline
\end{tabular}
\label{tab:temp_abund}
\end{table}

\begin{figure}
\centering
\includegraphics[width=16cm]{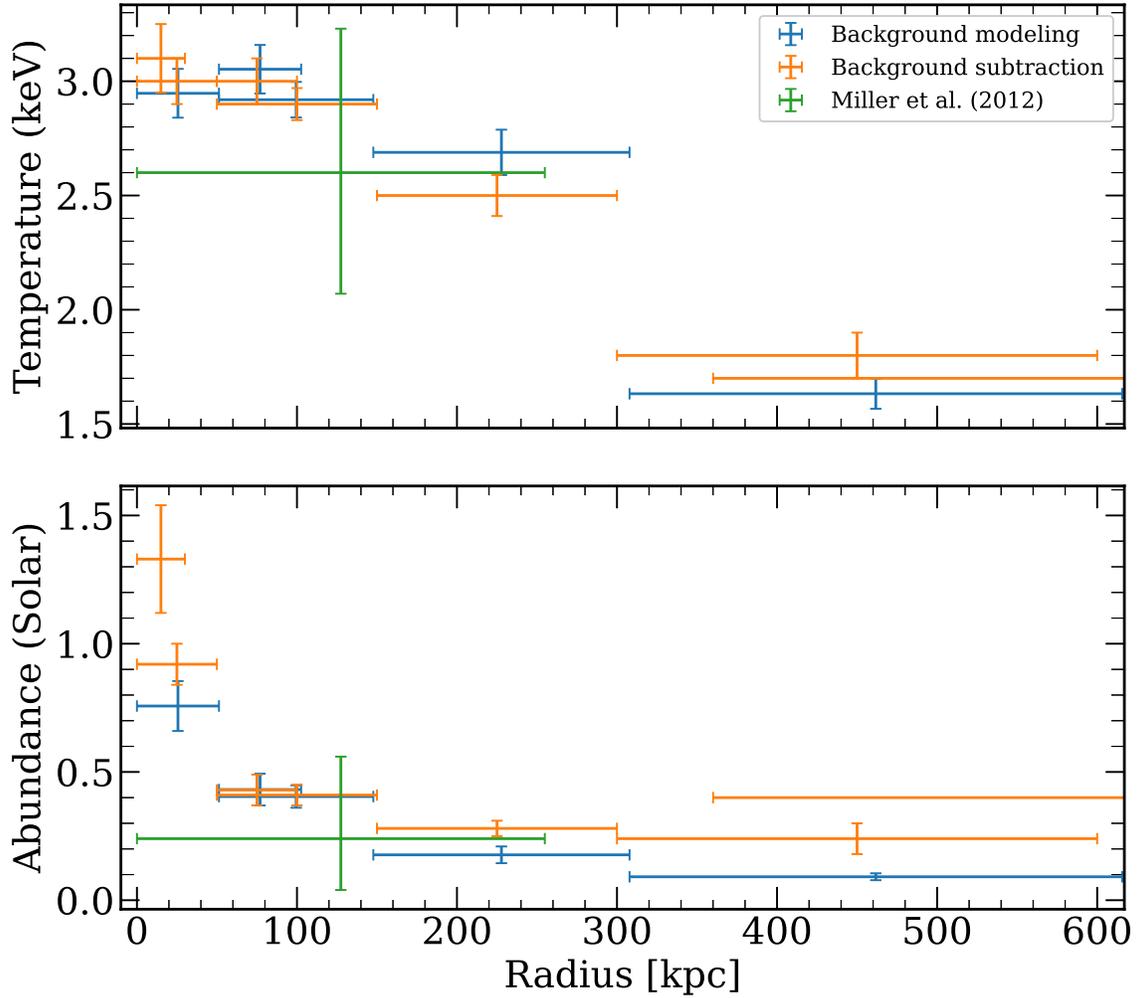}
\caption{Temperature (in keV) and metal abundance (in solar photospheric) radial profiles for \rxj~using a differential background modelling and an offset pointing.
We also show the best-fit values from Figure 5 for comparison. In green we show the values derived from a previous short Chandra observation \citep{eric-sample} for comparison.}
\label{fig:T-A-prof}
\end{figure}

To compare to the background treatment detailed in Section \ref{sec:bkg_treatment}, we performed an alternative protocol. Instead of subtracting the FWC events from each observed spectrum, we have modeled it simultaneously with the science observations \citep{2015A&A...575A..37M, 2017ApJ...851...69S,2019AJ....158....6S}. We include an additional offset pointing (PI:Koss, obsID 0821730401) found within 60$^{\prime}$ from the \rxj~ center. The EPIC observations were processed and filtered with the same tools described in Section \ref{sec:data_reduction}. Two types of background components are considered: the Astrophysical X-ray background (AXB) and non-X-ray background (NXB). The AXB model describes the emission of astrophysical sources, including the Local Bubble ({\tt apec$_{\rm LB}$}), hot Galactic Halo ({\tt apec$_{\rm MW}$}), and Cosmic X-ray background ({\tt powerlaw$_{\rm CXB}$}), as detailed in Section \ref{sec:bkg_treatment}. Abundance and redshift were fixed at 1 and 0, respectively, for both {\tt apec$_{\rm LB}$} and {\tt apec$_{\rm MW}$} models. We determine the AXB parameters by simultaneously fitting the offset pointing spectra from a circle radius of 12$^{\prime}$, its corresponding FWC spectra, and a RASS spectrum from a region of 0.3$^{\circ}$--0.9$^{\circ}$ ($\approx 2R_{200}-4R_{200}$) annulus, using the X-Ray Background Tool\footnote{https://heasarc.gsfc.nasa.gov/cgi-bin/Tools/xraybg/xraybg.pl}. The best-fit parameters for these components are presented in Table \ref{tab:bkg_params}.

The non-X-ray background (NXB) component describes the high energy particles that hit the CCD detectors during the observation. In the NXB model, each MOS and PN detector is represented by a set of fluorescent instrumental lines and a continuum spectrum.
Each instrumental line is modeled with a Gaussian model, where the line width is limited by $\leq0.3$\,keV, and the set of fluorescent lines considered are listed in Table 2 from \citet{2017ApJ...851...69S}. We also include the SWCX lines mentioned in Section \ref{sec:bkg_treatment}. Besides these lines, we model the continuum particle background with a broken power law, where the energy break is fixed at 3\,keV.
For each detector and observation, quiescent particle background data were generated using the {\tt evqpb} task. The event lists were filtered and cleaned with the same good time intervals, PATTERN, and FLAG as the observed data. To determine the NXB parameters, we first simultaneously fit the spectra of the unexposed corner data of both \rxj~ and its FWC (from the {\tt evqpb} task) for each EPIC observation to determine the ratios of their broken power law, which is fixed for each region of interest. A quiescent continuum of soft proton may persist even after filtering solar flare events. For each detector and observation, we compare the ratio of the count rates in the 6--12\,keV energy range from an inner region (removing 10$^\prime$) to those from the unexposed corner. To characterize the residual soft proton contamination, we add a power law if that ratio is above 1.15 \citep{2004A&A...419..837D}. The photon index of power law can vary between 0.1 and 1.4 \citep{Snowden14}.

The spectral fitting was performed in the 0.5--10.0\,keV and 0.7--10.0\,keV for MOS and PN detectors, respectively. For each region of interest, we jointly fit the \rxj~ and its FWC spectra for each observation. We consider the following set of models: {\tt wabs $\times$ (apec$_{\rm RXJ1007}$ + powerlaw$_{\rm CXB}$ + apec$_{\rm MW}$) + apec$_{\rm LB}$}. All FWC data were set to zero for this model. The AXB parameters were fixed at the best-fit values listed in Table \ref{tab:bkg_params}. The Galactic extinction is modeled through the {\tt wabs} model, which component was fixed to the reported average nH value of 1.36$\times$10$^{20}$cm$^{-2}$ \citep{HI4PI}. The emission of the \rxj~ hot gas, {\tt apec$_{\rm RXJ1007}$}, is modeled for each region of interest with redshift fixed at 0.112; the plasma temperature, abundance, and emission intensity (normalization) were free to vary. The NXB model contains the NXB components, including the broken power law, set of fluorescent instrumental lines, and possibly the power law to characterize the quiescent continuum of the soft proton. We link the NXB parameters from the observed spectra and its FWC data for each region of interest. We performed all spectral fittings using XSPEC {\tt v12.12} and $\chi^2$ statistics. We report the error-weighted average abundance and temperature with a 1$\sigma$ confidence level by implementing joint fit of the MOS1, MOS2, and PN detectors fr each observation in Figure \ref{fig:T-A-prof}.

\section{Catalog of galaxies with confirmed spectroscopic redshifts} \label{appx:galaxy_redshift}

Table~\ref{tab:galmem} lists the member galaxies of \rxjfull~inside a radius of $\sim$ 12\arcmin~($\sim$ 1.4 Mpc) obtained with GMOS and with SDSS DR15. Table~\ref{tab:galnomem} shows the list of galaxies observed with GMOS and located in the foreground and background of the \rxjfull.

\begin{deluxetable}{cccccccccc}
\tabletypesize{\scriptsize}
\tablecolumns{10}
\tablecaption{Catalog of member galaxies with confirmed spectroscopic redshifts\label{tab:galmem}}
\tablehead{
\colhead{Galaxy} & \colhead{Obj. ID} & \colhead{Ref.} & \colhead{R.A.} & \colhead{Decl.} & \colhead{\sloang} & \colhead{\sloanr} & 
\colhead{\sloangr} & \colhead{z $\pm$ $\Delta z$} & \colhead{R\#l}\\
\colhead{(1)} & \colhead{(2)} & \colhead{(3)} & \colhead{(4)} & \colhead{(5)} & \colhead{(6)} & \colhead{(7)} &
\colhead{(8)} & \colhead{(9)} & \colhead{(10)}}
\startdata
J100652.19+375847.6 & 558798 & SDSS & 151.7174740391 & $+$37.9799012130 & 18.11 & 17.35 & 0.76 & 0.112662$\pm$0.000018 &\nodata/\nodata \\ 
J100657.90+381125.4 & 129132 & SDSS & 151.7412550403 & $+$38.1904025627 & 18.43 & 17.43 & 1.00 & 0.113578$\pm$0.000028 &\nodata/\nodata \\ 
J100715.06+380635.5 & 624446 & SDSS & 151.8127399514 & $+$38.1098518985 & 18.25 & 17.30 & 0.95 & 0.110774$\pm$0.000028 &\nodata/\nodata \\ 
J100716.55+375915.5 & 624393 & SDSS & 151.8189433735 & $+$37.9876400048 & 17.18 & 16.23 & 0.95 & 0.111714$\pm$0.000021 &\nodata/\nodata \\ 
J100719.95+380316.7 & 624439 & SDSS & 151.8331448606 & $+$38.0546491373 & 19.09 & 18.16 & 0.93 & 0.110354$\pm$0.000023 &\nodata/\nodata \\ 
J100720.90+375944.8 & 624416 & SDSS & 151.8371017052 & $+$37.9957726440 & 17.98 & 17.00 & 0.98 & 0.109762$\pm$0.000024 &\nodata/\nodata \\ 
J100726.52+375951.3 & 624435 & SDSS & 151.8605170752 & $+$37.9975862200 & 18.96 & 18.00 & 0.96 & 0.110808$\pm$0.000024 &\nodata/\nodata \\ 
J100731.38+380039.5 & 624450 & GMOS & 151.8807620208 & $+$38.0109700942 & 19.56 & 18.61 & 0.95 & 0.109506$\pm$0.000304 & 3.61/\nodata \\ 
J100733.54+380110.9 & 624670 & GMOS & 151.8897374004 & $+$38.0196820358 & 19.29 & 18.93 & 0.36 & 0.109639$\pm$0.000051 & \nodata/9 \\
J100736.74+380019.8 & 624264 & GMOS & 151.9030646984 & $+$38.0054931863 & 19.37 & 18.41 & 0.95 & 0.114143$\pm$0.000268 & 6.26/\nodata \\ 
J100737.25+380325.6 & 625056 & GMOS & 151.9051928292 & $+$38.0571049518 & 21.93 & 21.48 & 0.45 & 0.113242$\pm$0.000145 &\nodata/\nodata \\ 
J100737.37+380342.3 & 624474 & SDSS & 151.9057147685 & $+$38.0617362630 & 19.08 & 18.14 & 0.94 & 0.108442$\pm$0.000025 &\nodata/\nodata \\ 
J100737.60+380225.9 & 624469 & GMOS & 151.9066709091 & $+$38.0405296324 & 19.65 & 18.72 & 0.93 & 0.113026$\pm$0.000189 & 6.30/\nodata \\ 
J100737.87+375901.2 & 624553 & GMOS & 151.9078075029 & $+$37.9836546686 & 19.96 & 19.05 & 0.91 & 0.113752$\pm$0.000195 & 5.80/\nodata \\ 
J100738.57+375952.3 & 624265 & GMOS & 151.9106924003 & $+$37.9978683739 & 19.85 & 18.88 & 0.97 & 0.113459$\pm$0.000259 & 4.49/\nodata \\ 
J100740.78+375945.0 & 624262 & GMOS & 151.9199037132 & $+$37.9958265094 & 19.19 & 18.20 & 0.99 & 0.111000$\pm$0.000174 & 8.41/\nodata \\ 
J100741.94+375919.4 & 624250 & GMOS & 151.9247507026 & $+$37.9887201994 & 21.83 & 21.04 & 0.79 & 0.111795$\pm$0.000188 & 3.38/\nodata \\ 
J100742.24+380129.5 & 624478 & GMOS & 151.9260107138 & $+$38.0248726653 & 18.26 & 17.24 & 1.02 & 0.109932$\pm$0.000267 & 5.61/\nodata \\ 
J100742.38+375843.2 & 625019 & GMOS & 151.9265651695 & $+$37.9786694145 & 22.22 & 21.42 & 0.80 & 0.109236$\pm$0.000156 & 4.69/\nodata \\ 
J100742.53+380046.6 & 624259 & GMOS & 151.9272024039 & $+$38.0129438025 & 15.79 & 14.73 & 1.06 & 0.112111$\pm$0.000162 & 10.43/\nodata \\ 
J100744.23+380025.5 & 624263 & GMOS & 151.9342861715 & $+$38.0070925323 & 19.24 & 18.31 & 0.93 & 0.108319$\pm$0.000212 & 5.33/\nodata \\ 
J100744.40+375916.5 & 624471 & GMOS & 151.9349819179 & $+$37.9879053184 & 18.79 & 17.82 & 0.97 & 0.117360$\pm$0.000127 & 10.52/\nodata \\ 
J100744.40+380115.6 & 624261 & GMOS & 151.9350060978 & $+$38.0209965538 & 19.48 & 18.40 & 1.09 & 0.114679$\pm$0.000317 & 3.92/\nodata \\ 
J100744.65+380134.5 & 624483 & GMOS & 151.9360298506 & $+$38.0262489794 & 19.51 & 18.51 & 1.01 & 0.112577$\pm$0.000131 & 8.33/\nodata \\ 
J100746.61+380224.9 & 624500 & GMOS & 151.9442059323 & $+$38.0402503045 & 19.15 & 18.16 & 0.98 & 0.113174$\pm$0.000174 & 5.43/\nodata \\ 
J100747.79+380520.0 & 689797 & SDSS & 151.9491165793 & $+$38.0888772164 & 17.96 & 17.24 & 0.72 & 0.110905$\pm$0.000011 &\nodata/\nodata \\ 
J100750.00+380302.8 & 689932 & GMOS & 151.9583347008 & $+$38.0507903425 & 20.58 & 19.68 & 0.90 & 0.113259$\pm$0.000167 & 6.25/\nodata \\ 
J100750.53+375638.3 & 258143 & SDSS & 151.9605475467 & $+$37.9439595124 & 18.49 & 17.59 & 0.91 & 0.111268$\pm$0.000026 &\nodata/\nodata \\ 
J100751.59+380206.7 & 689782 & GMOS & 151.9649661918 & $+$38.0352039618 & 18.43 & 17.49 & 0.95 & 0.109543$\pm$0.000233 & 6.78/\nodata \\ 
J100753.84+380008.0 & 624704 & GMOS & 151.9743297133 & $+$38.0022127342 & 21.07 & 20.97 & 0.10 & 0.111607$\pm$0.000115 &\nodata/\nodata \\ 
J100758.99+375427.7 & 258239 & SDSS & 151.9957983609 & $+$37.9076823814 & 19.04 & 18.02 & 1.02 & 0.114132$\pm$0.000020 &\nodata/\nodata \\ 
J100759.70+380016.0 & 689803 & SDSS & 151.9987339394 & $+$38.0044310006 & 18.49 & 17.67 & 0.82 & 0.107024$\pm$0.000027 &\nodata/\nodata \\ 
J100802.51+380025.3 & 689809 & SDSS & 152.0104583039 & $+$38.0070299295 & 18.10 & 17.14 & 0.96 & 0.107167$\pm$0.000018 &\nodata/\nodata \\ 
J100804.46+375554.1 & 323638 & SDSS & 152.0186012978 & $+$37.9316814362 & 19.49 & 18.56 & 0.93 & 0.112875$\pm$0.000028 &\nodata/\nodata \\ 
J100807.65+375606.4 & 323658 & SDSS & 152.0318840949 & $+$37.9351225528 & 17.70 & 16.73 & 0.97 & 0.113735$\pm$0.000034 &\nodata/\nodata \\ 
J100808.40+380011.2 & 323775 & SDSS & 152.0349998027 & $+$38.0031001111 & 17.76 & 17.01 & 0.74 & 0.109603$\pm$0.000046 &\nodata/\nodata \\ 
J100808.99+375853.8 & 323702 & SDSS & 152.0374528327 & $+$37.9816057300 & 18.62 & 17.63 & 0.99 & 0.113330$\pm$0.000027 &\nodata/\nodata \\ 
J100809.65+380211.5 & 689825 & SDSS & 152.0402019167 & $+$38.0365389152 & 18.99 & 18.02 & 0.96 & 0.110790$\pm$0.000026 &\nodata/\nodata \\ 
J100814.25+380349.9 & 689844 & SDSS & 152.0593782040 & $+$38.0638552695 & 18.63 & 17.60 & 1.03 & 0.112135$\pm$0.000031 &\nodata/\nodata \\ 
J100816.26+375749.6 & 323706 & SDSS & 152.0677364064 & $+$37.9637703098 & 18.99 & 18.05 & 0.94 & 0.110897$\pm$0.000027 &\nodata/\nodata \\ 
J100818.65+380457.9 & 689868 & SDSS & 152.0777099713 & $+$38.0827616593 & 17.38 & 16.33 & 1.06 & 0.112165$\pm$0.000023 &\nodata/\nodata \\ 
J100820.17+380204.0 & 689855 & SDSS & 152.0840308341 & $+$38.0344577558 & 18.32 & 17.30 & 1.01 & 0.111328$\pm$0.000029 &\nodata/\nodata \\ 
J100827.09+380322.6 & 689877 & SDSS & 152.1128733163 & $+$38.0562855840 & 18.58 & 17.60 & 0.98 & 0.112722$\pm$0.000032 &\nodata/\nodata \\ 
J100827.23+380503.6 & 689897 & SDSS & 152.1134508266 & $+$38.0843458319 & 18.59 & 17.70 & 0.89 & 0.113091$\pm$0.000027 &\nodata/\nodata \\ 
J100830.85+375907.5 & 323731 & SDSS & 152.1285249209 & $+$37.9854127103 & 17.39 & 16.59 & 0.80 & 0.114512$\pm$0.000025 &\nodata/\nodata \\ 
J100836.32+375445.5 & 323723 & SDSS & 152.1513383034 & $+$37.9126505738 & 17.87 & 17.16 & 0.70 & 0.110156$\pm$0.000043 &\nodata/\nodata \\ 
\enddata
\tablecomments{The meaning of the columns are the following: (1) - Galaxy name; (2) - SDSS DR15 object id; (3) - Redshift source; 
(4) - (5) - Right Ascension and Declination (J2000.0) in units of degrees; (6) and (7) - \sloang and \sloanr~model magnitudes; 
(8) - \sloangr~color ; (9) -  galaxy redshifts and associated errors: (10) - $R$ values (Tonry \& Davis 1979 - real numbers) or 
the number of emission lines (integer values) used to calculate redshifts}
=\end{deluxetable}

\begin{deluxetable}{ccccccccc}
=\tabletypesize{\scriptsize}
\tablecolumns{9}
\tablecaption{Catalog of  foreground and background galaxies observed with GMOS\label{tab:galnomem}}
\tablehead{
\colhead{Galaxy} & \colhead{Obj. ID} & \colhead{R.A.} & \colhead{Decl.} & \colhead{\sloang} & \colhead{\sloanr} & 
\colhead{\sloangr} & \colhead{z $\pm$ $\Delta z$} & \colhead{R\#l}\\
\colhead{(1)} & \colhead{(2)} & \colhead{(3)} & \colhead{(4)} & \colhead{(5)} & \colhead{(6)} & \colhead{(7)} &
\colhead{(8)} & \colhead{(9)}}
\startdata
J100733.92+375942.0 & 624955 & 151.8913532072 & $+$37.9950010354 & 21.30 & 20.68 & 0.62 & 0.145270$\pm$0.000127 & \nodata/11 \\
J100734.06+380102.3 & 624979 & 151.8919164309 & $+$38.0172934724 & 20.82 & 19.98 & 0.84 & 0.124720$\pm$0.000127 & 4.98/\nodata \\
J100734.47+375841.9 & 624942 & 151.8936185352 & $+$37.9783133370 & 21.53 & 20.68 & 0.85 & 0.121544$\pm$0.000245 & 4.04/\nodata \\
J100734.65+380137.4 & 624459 & 151.8943944995 & $+$38.0270678393 & 18.04 & 17.58 & 0.46 & 0.051796$\pm$0.000049 & \nodata/9 \\
J100737.36+380217.8 & 625037 & 151.9056463422 & $+$38.0382682727 & 22.12 & 21.31 & 0.81 & 0.076991$\pm$0.000288 & 2.48/\nodata \\
J100739.34+380118.3 & 625035 & 151.9139174207 & $+$38.0217458542 & 21.67 & 20.24 & 1.43 & 0.463894$\pm$0.000132 & 8.29/\nodata \\
J100739.67+380241.4 & 624687 & 151.9152762652 & $+$38.0448398176 & 20.99 & 20.27 & 0.72 & 0.449444$\pm$0.000088 & \nodata/7 \\
J100740.17+380007.6\tablenotemark{a} & 900002 & 151.9173750000 & $+$38.0021111100 &\nodata &\nodata  &\nodata& 0.458301$\pm$0.000120 & 10.10/\nodata \\
J100740.43+380029.9 & 624280 & 151.9184638520 & $+$38.0083110006 & 22.40 & 21.40 & 1.00 & 0.386789$\pm$0.000167 & 3.95/\nodata \\
J100740.77+380223.3 & 625078 & 151.9198880387 & $+$38.0397946483 & 21.38 & 19.76 & 1.62 & 0.326552$\pm$0.000192 & 6.76/\nodata \\
J100741.34+380331.0 & 625098 & 151.9222365169 & $+$38.0586229403 & 22.78 & 21.14 & 1.64 & 0.131661$\pm$0.000279 & 3.16/\nodata \\
J100741.36+380104.6 & 624266 & 151.9223377080 & $+$38.0179565629 & 19.57 & 18.11 & 1.46 & 0.157989$\pm$0.000097 & 7.81/\nodata \\
J100742.57+380240.6 & 625095 & 151.9273874324 & $+$38.0446071322 & 22.49 & 21.46 & 1.02 & 0.411458$\pm$0.000152 & \nodata/7 \\
J100743.22+380107.1 & 624275 & 151.9300727167 & $+$38.0186453694 & 21.82 & 20.90 & 0.92 & 0.442847$\pm$0.000194 & 3.12/\nodata \\
J100743.65+375954.0 & 625048 & 151.9318769142 & $+$37.9983358319 & 22.23 & 21.21 & 1.01 & 0.138400$\pm$0.000162 & 4.82/\nodata \\
J100743.76+380328.9 & 625131 & 151.9323460813 & $+$38.0580247540 & 22.18 & 21.40 & 0.78 & 0.383729$\pm$0.000332 & 2.14/\nodata \\
J100743.80+380057.2 & 624274 & 151.9325140533 & $+$38.0158864819 & 22.13 & 21.01 & 1.12 & 0.348508$\pm$0.000377 & 2.71/\nodata \\
J100744.20+375833.5 & 624597 & 151.9341701239 & $+$37.9759837465 & 21.65 & 20.87 & 0.78 & 0.543780$\pm$0.000116 & \nodata/7 \\
J100744.32+375800.9 & 258348 & 151.9346642278 & $+$37.9669038069 & 20.41 & 18.85 & 1.57 & 0.384531$\pm$0.000173 & 7.77/\nodata \\
J100744.90+380320.2 & 625134 & 151.9370929375 & $+$38.0556018121 & 22.11 & 21.37 & 0.74 & 0.181325$\pm$0.000276 & 3.03/\nodata \\
J100744.93+375927.2 & 624685 & 151.9372123404 & $+$37.9908771878 & 21.33 & 20.35 & 0.98 & 0.450612$\pm$0.000402 & \nodata/6 \\
J100745.71+375828.4 & 258618 & 151.9404527460 & $+$37.9745664004 & 21.01 & 19.37 & 1.64 & 0.327555$\pm$0.000142 & 10.12/\nodata \\
J100745.98+375855.1 & 625051 & 151.9415879796 & $+$37.9819847401 & 22.62 & 21.48 & 1.14 & 0.383415$\pm$0.000272 & \nodata/8 \\
J100746.49+375834.7\tablenotemark{a} & 900001 & 151.9437083300 & $+$37.9763055600 &\nodata &\nodata  &\nodata& 0.383022$\pm$0.000079 & \nodata/8 \\
J100747.11+375808.1 & 258645 & 151.9463048288 & $+$37.9689267981 & 21.62 & 20.46 & 1.16 & 0.445158$\pm$0.000320 & \nodata/6 \\
J100747.29+380155.5 & 900001 & 151.9470705420 & $+$38.0320911220 & 21.46 & 20.45 & 1.02 & 0.213481$\pm$0.000110 & \nodata/12 \\
J100747.30+380157.8 & 624496 & 151.9470911999 & $+$38.0327131903 & 19.00 & 17.90 & 1.10 & 0.212334$\pm$0.000272 & 3.29/\nodata \\
J100747.33+380150.5 & 624498 & 151.9472065144 & $+$38.0306955557 & 20.56 & 19.93 & 0.63 & 0.570161$\pm$0.000225 & 3.14/\nodata \\
J100747.42+380300.6 & 624510 & 151.9476013619 & $+$38.0501701275 & 21.96 & 21.22 & 0.74 & 0.545821$\pm$0.000139 & \nodata/8 \\
J100747.98+380215.5 & 624701 & 151.9499175795 & $+$38.0376375166 & 20.72 & 19.72 & 1.00 & 0.149255$\pm$0.000136 & 3.97/\nodata \\
J100748.81+375800.1 & 258660 & 151.9533828082 & $+$37.9666857690 & 21.56 & 20.54 & 1.02 & 0.213040$\pm$0.000242 & 4.83/\nodata \\
J100748.82+380231.9 & 624575 & 151.9534116498 & $+$38.0421944386 & 20.93 & 19.44 & 1.48 & 0.274047$\pm$0.000184 & 5.16/\nodata \\
J100748.83+380243.7 & 625153 & 151.9534553928 & $+$38.0454681629 & 20.32 & 19.25 & 1.06 & 0.138699$\pm$0.000208 & 3.47/\nodata \\
J100748.85+380127.1 & 625136 & 151.9535556120 & $+$38.0241871473 & 20.03 & 19.07 & 0.96 & 0.211376$\pm$0.000168 & \nodata/10 \\
J100749.63+375958.4 & 625116 & 151.9568002960 & $+$37.9995614266 & 22.59 & 21.43 & 1.17 & 0.326353$\pm$0.000193 & 3.17/\nodata \\
J100749.70+380114.4 & 624566 & 151.9570669979 & $+$38.0206611312 & 20.12 & 19.36 & 0.75 & 0.212394$\pm$0.000237 & \nodata/10 \\
J100749.82+375807.8 & 258669 & 151.9575748909 & $+$37.9688362186 & 20.53 & 19.81 & 0.72 & 0.161819$\pm$0.000260 & \nodata/8 \\
J100750.15+375747.3 & 258667 & 151.9589718626 & $+$37.9631437189 & 22.38 & 21.43 & 0.94 & 0.457767$\pm$0.000259 & \nodata/6 \\
J100750.40+375931.3 & 624561 & 151.9599971366 & $+$37.9920139289 & 20.89 & 19.27 & 1.61 & 0.327404$\pm$0.000283 & 5.60/\nodata \\
J100751.38+375837.1 & 258698 & 151.9640873401 & $+$37.9769667988 & 21.37 & 19.86 & 1.51 & 0.326841$\pm$0.000295 & 4.04/\nodata \\
J100751.67+380212.3 & 689783 & 151.9652953858 & $+$38.0367425656 & 22.33 & 21.34 & 0.99 & 0.406271$\pm$0.000332 & \nodata/7 \\
J100751.75+380240.0 & 690000 & 151.9656325346 & $+$38.0444460621 & 21.16 & 20.45 & 0.71 & 0.275814$\pm$0.000241 & \nodata/8 \\
J100752.19+380127.7 & 690114 & 151.9674533681 & $+$38.0243658239 & 21.29 & 20.46 & 0.83 & 0.177286$\pm$0.000220 & 3.79/\nodata \\
J100752.85+380243.1 & 690145 & 151.9702251654 & $+$38.0453168391 & 21.71 & 20.78 & 0.93 & 0.073080$\pm$0.000155 & 5.10/\nodata \\
J100753.13+380248.8 & 690146 & 151.9713544637 & $+$38.0468871037 & 21.29 & 20.27 & 1.03 & 0.094176$\pm$0.000265 & 3.30/\nodata \\
J100754.00+375912.8 & 625142 & 151.9750105213 & $+$37.9868776082 & 21.46 & 19.91 & 1.54 & 0.327918$\pm$0.000146 & 8.25/\nodata \\
\enddata
\tablecomments{The meaning of the columns are the following: (1) - Galaxy name; (2) - SDSS DR15 object id;
(3) - (4) - Right Ascension and Declination (J2000.0) in units of degrees; (5) and (6) - \sloang and \sloanr~model magnitudes; 
(7) - \sloangr~color ; (8) -  galaxy redshifts and associated errors: (9) - $R$ values (Tonry \& Davis 1979 - real numbers) or 
the number of emission lines (integer values) used to calculate redshifts}
=\tablenotetext{a}{Galaxies with no magnitude information in the SDSS DR15 database}
\end{deluxetable}

\bibliography{RXJ1007.bib}
\bibliographystyle{aasjournal}

\end{document}